\documentclass[aps,prb,twocolumn,superscriptaddress,showpacs]{revtex4}
\usepackage{mathtools,amsmath,amsfonts,amssymb,graphicx,color,bm, soul, color, bbold,stmaryrd,wasysym}
\usepackage{verbatim}

\begin{document}
\title{Characterizing the rate and coherence of single-electron tunneling between two dangling bonds on the surface of silicon}
\author{Zahra Shaterzadeh-Yazdi}
\email{zshaterz@ucalgary.ca}
\affiliation{Institute for Quantum Science and Technology, University of Calgary, Alberta T2N 1N4, Canada}
\author{Lucian Livadaru }
\affiliation{National Institute for Nanotechnology, National Research Council of Canada, Edmonton, Alberta T6G 2M9, Canada}
\affiliation{Department of Physics, University of Alberta, Edmonton, Alberta T6G 2J1, Canada}
\author{Marco Taucer}
\affiliation{National Institute for Nanotechnology, National Research Council of Canada, Edmonton, Alberta T6G 2M9, Canada}
\affiliation{Department of Physics, University of Alberta, Edmonton, Alberta T6G 2J1, Canada}
\author{Josh Mutus}
\affiliation{National Institute for Nanotechnology, National Research Council of Canada, Edmonton, Alberta T6G 2M9, Canada}
\affiliation{Department of Physics, University of Alberta, Edmonton, Alberta T6G 2J1, Canada}

\author{Jason Pitters}
\affiliation{National Institute for Nanotechnology, National Research Council of Canada, Edmonton, Alberta T6G 2M9, Canada}
\affiliation{Department of Physics, University of Alberta, Edmonton, Alberta T6G 2J1, Canada}
\author{Robert A. Wolkow}
\affiliation{National Institute for Nanotechnology, National Research Council of Canada, Edmonton, Alberta T6G 2M9, Canada}
\affiliation{Department of Physics, University of Alberta, Edmonton, Alberta T6G 2J1, Canada}
\author{Barry C. Sanders}
\affiliation{Institute for Quantum Science and Technology, University of Calgary, Alberta T2N 1N4, Canada}
\date{\today}

\begin{abstract}
We devise a scheme to characterize tunneling of an excess electron shared 
by a pair of tunnel-coupled dangling-bonds on a silicon surface -- effectively a two-level system.
Theoretical estimates show that the tunneling should be highly coherent but too fast 
to be measured by any conventional techniques.
Our approach is instead to measure the time-averaged charge distribution 
of our dangling-bond pair 
by a capacitively coupled atomic-force-microscope tip in the presence of both 
a surface-parallel electrostatic potential bias between the two dangling bonds and 
a tunable mid-infrared laser capable of inducing Rabi oscillations in the system.
With a nonresonant laser, the time-averaged charge distribution in the dangling-bond pair
is asymmetric as imposed by the bias. 
However, as the laser becomes resonant 
with the coherent electron tunneling in the biased  pair
the theory predicts that  
the time-averaged charge distribution becomes symmetric.
This resonant symmetry effect should not only reveal the tunneling rate, 
but also the nature and rate of decoherence of single electron dynamics in our system.
\end{abstract}
\maketitle
\section{Introduction}
\label{sec:intro}

Two closely spaced quantum dots can function as an effective two-level system
when they share an extra electron via coherent tunneling.
Such a system is also known as a ``charge qubit'' and its
various physical implementations have been explored
with applications to quantum computing,\cite{NPT99, HDW+04, HFC+03} 
spin-charge conversion~\cite{HFC+03, GHW05} for spin-qubit readout,\cite{EHW+04}
and in general for engineering new devices on silicon surfaces at the quantum level.\cite{schofield2013quantum}
Although solid-state charge qubits have been demonstrated
in semiconductors and superconductors,\cite{NPT99, WSB+05}
in practice their properties exhibit uncontrolled and undesired 
variability (of the single dot and of the entire assembly) 
due to growth imperfections and decoherence mechanisms.

We have suggested overcoming these challenges by employing silicon-surface dangling bonds (DBs) as 
tunnel-coupled quantum dots sharing a controllable number of electrons~\cite{HPD+09}.
All such dots are identical and their spacing is determined with atomic-scale precision.
An excess electron shared in a pair of such DBs is predicted to be highly coherent 
with a tunneling period between 10~fs and 1~ps.\cite{LXS+10, PLH+11}
Furthermore, the fabrication of relatively large assemblies of DBs 
can be achieved by using a scanning tunneling microscope with
great precision, reliability, reproducibility, and virtually no variability at the single dot level.

However, the very same feature that leads to high coherence,
namely  the great rate of tunneling, also gives rise to
some considerable practical difficulties in that direct characterization
by monitoring the oscillation is not feasible electronically by any straightforward methods.
Here we propose a strategy to measure the rate and coherence of tunneling by controlling 
and monitoring time-averaged charge distributions in pairs of coupled DB dots.\cite{GBC11}
These measurements are inspired by previous experiments on double quantum-dot 
structures with tunneling rates in the microwave regime.\cite{PJM+04, SDB05}

A DB pair can be created such that electrostatic repulsion prevents 
an excess electron acquired at each DB (i.e. $-2e$ net charge per pair),
and DBs are sufficiently close so that they are tunnel-coupled with each other.
The DB pair can thus be geometrically tuned to have the desired 
occupation of one extra electron per pair, and henceforth we call it DBP$^-$ for convenience.
In our approach, the electron's position within the `left' (L) or `right' (R) dot 
is discerned by an atomic force microscope (AFM) capacitively coupled to the pair.

Atomic force microscopy can achieve single-electron sensitivity\cite{GBC11, ZBM05} and recently it has been used for detecting the electronic properties of individual and coupled quantum dots in contact with a reservoir.\cite{CMB+10, CMB+12}
A tunnel-coupled DB system is analogous to these coupled quantum dot system, but on a smaller spatial scale. 
As such, they exhibit molecular state formations as in quantum dot systems or in 
``natural" molecules. 
Similar to other quantum dot systems, artificially fabricated DB bonds offer the 
possibility of choosing bond strength  (tunnel splitting) via their geometry.
Unlike conventional chemical bonds, the through-space bonding occurs partly in vacuum and partly in the silicon dielectric medium.
Such a particular flavor of bond will no doubt be the subject of investigations beyond the scope of this study.

An alternative but equivalent picture is that of coupled DB pairs with electrons 
tunneling coherently between individual DBs.
As the AFM measurements are in practice relatively slow on the time scale of 
the DB pair charge dynamics,
they average over many oscillations thereby losing all direct 
information about the tunneling rate and decoherence.
Effectively, during such a measurement, capacitive coupling to the DBP$^-$
induces an anharmonic component in the potential on the AFM tip, which is
otherwise harmonic in the absence of capacitive coupling to localized charges.\cite{Gie03}
Importantly, the strength of this coupling reveals the time-averaged charge in the 
left (and right) dot and is manifested as a shift in the AFM oscillation frequency. 

By using lithographic contacts on a silicon surface, it is generally 
possible to apply electric potential biases in order to
locally address single atoms and molecules.\cite{FMM+12, ZDS+11, PDW11}
Such contacts can be used to establish an electric field along the two DBs in a DBP$^-$.
This bias will cause the probability distribution for the position of an excess electron 
to be more heavily weighted in the left or right DB depending on the sign and strength of the bias. 
In our scheme it is exactly this distribution that is observable by AFM.
Furthermore, the actual tunneling rate is influenced by this static bias.
For zero bias the position of the excess electron is equally probable in the right and left DB.

In order to experimentally determine tunneling rate and decoherence, another ingredient is needed:
an oscillatory driving force pushing the excess electron back and forth rapidly between the two DBs
at a rate comparable to the native tunneling frequency of the DBP$^-$.
In the case of two DBs, the driving field needs to be in the mid-infrared (MIR) regime.

If the MIR field is off-resonant with the inter-dot tunneling frequency,
the resultant force has only a small perturbative effect on the double-dot system
so that the excess electron distribution is nearly the same as that without a driving field.
If the MIR radiation is resonant, it can be theoretically shown that its field causes the electron 
to be equally probable at either DB.
As one varies the frequency of the driving field, the above effect results in a relatively 
abrupt change in the excess electron distribution and causes an equally abrupt change 
in the AFM tip-oscillation frequency, which can be measured experimentally.
The response of the AFM tip should reveal the tunneling rate 
and some properties and parameters of decoherence.
The remainder of this paper elaborates on this concept and provides the technical details of our approach. 

\section{Background}
\label{sec:background}
\begin{figure}
	\includegraphics[width=0.8\columnwidth]{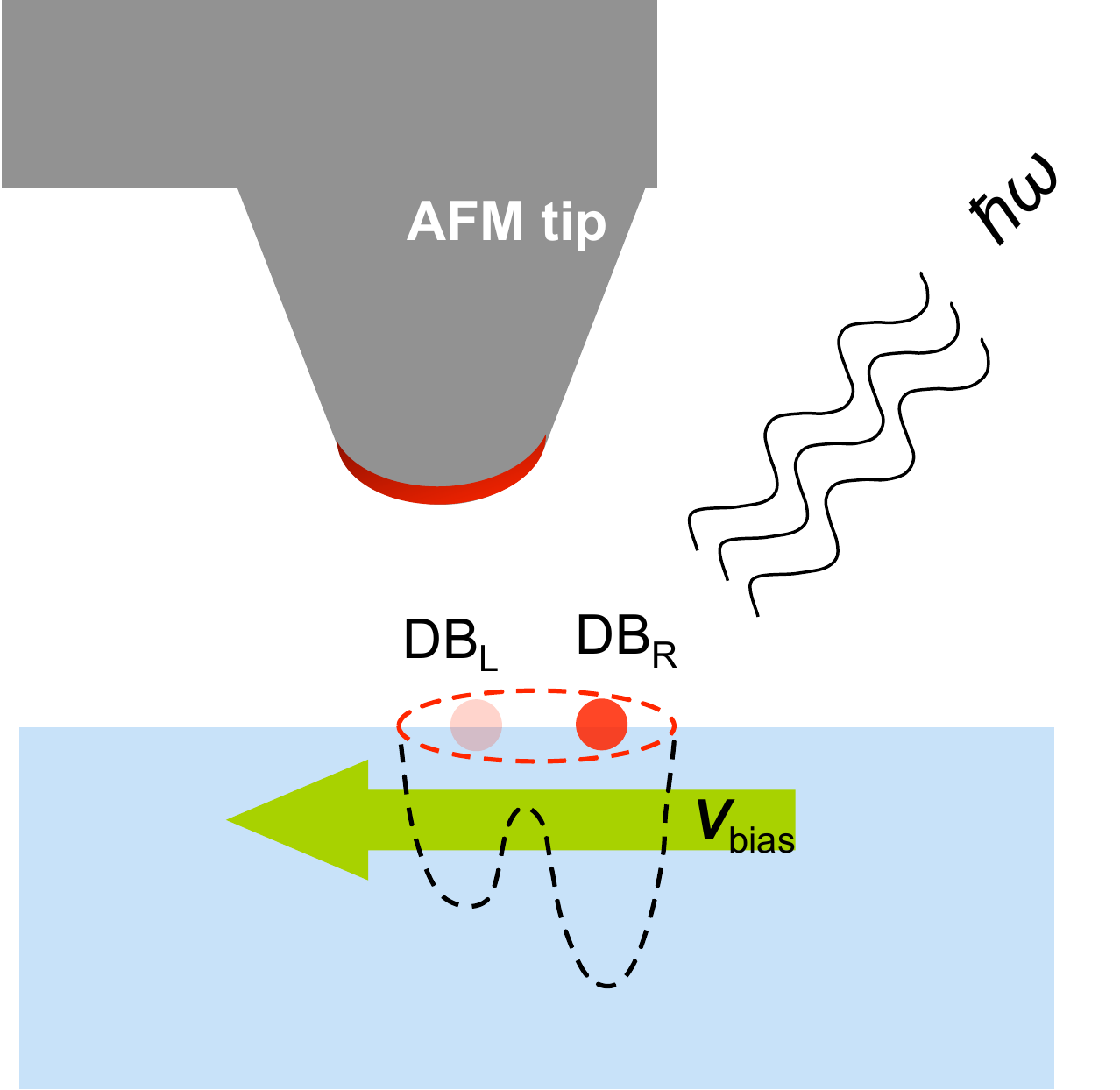}
\caption{
	Dangling-bond pair (DB$_\text{L}$-DB$_\text{R}$)
	is depicted as a double-well potential at the silicon-vacuum interface. 
    An excess electron shown 
    as a red (dark) dot oscillates between the two wells.
	The DB pair is subjected to a static electric bias 
	and driven by laser radiation.
	An atomic-force microscope (AFM) tip is capacitively coupled to the DB pair due to electrostatic 
	interaction between charges on the AFM tip (red (dark) zone on tip apex) and the excess  
	electron in the double-well potential. The AFM tip oscillates
	with a frequency that is dependent on the location of this excess charge
	thereby modifying the tip oscillation frequency in a predictable way. 
	}
\label{fig:N1scheme}
\end{figure}

The scheme discussed in Sec.~\ref{sec:intro} is shown in Fig.~\ref{fig:N1scheme} 
and described in detail in the figure caption.
The scheme comprises a DBP$^-$ subject to a static potential bias, an AFM tip capacitively 
coupled to the DBs, and a MIR driving field.
In this section we discuss technical issues concerning each component of the proposed apparatus.

\subsection{Dangling bond on H--Si(100)--2$\times$1}
\label{subsec:coupled}

\begin{figure}
	\includegraphics[width=\columnwidth]{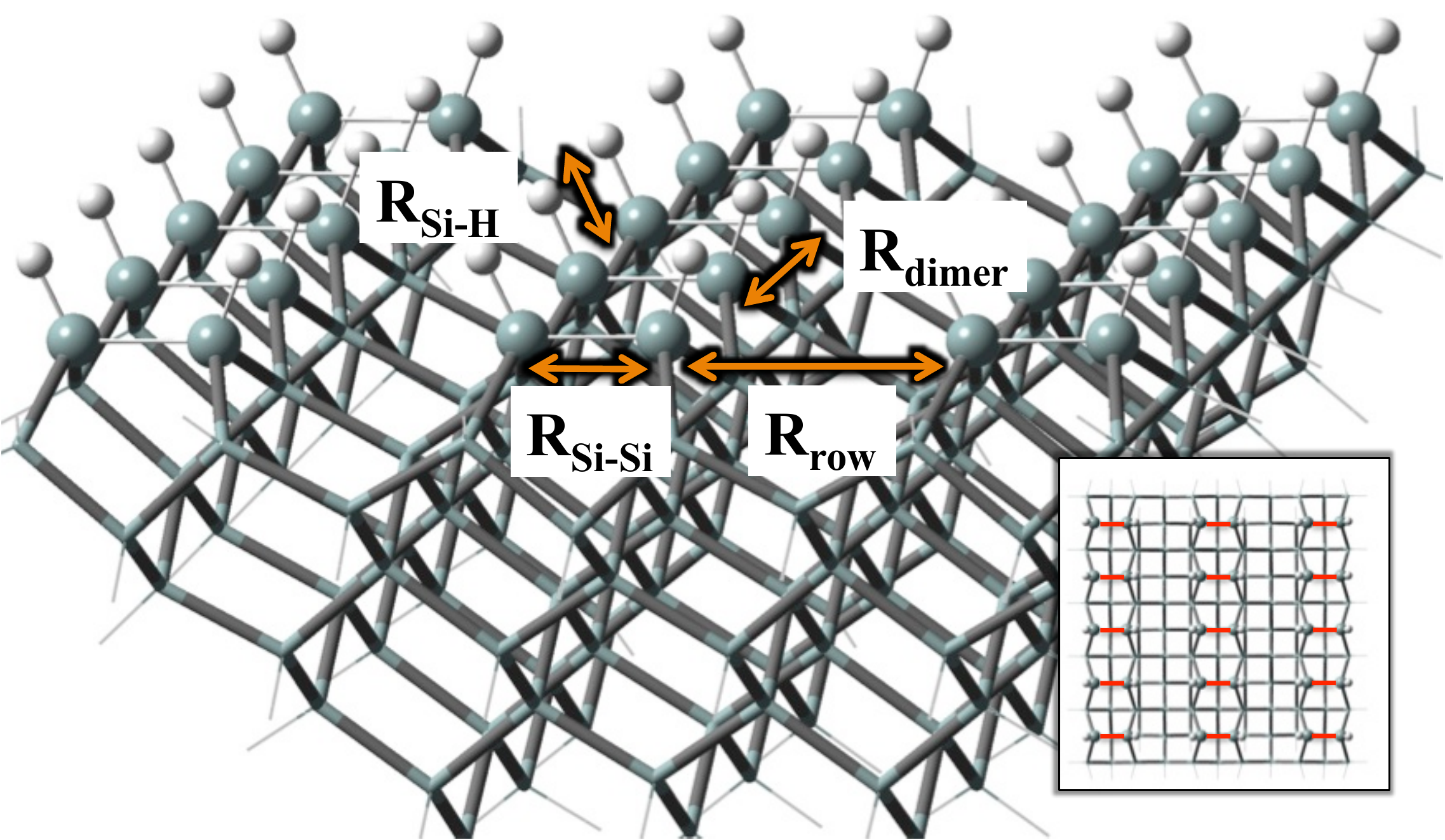}
\caption{
	Surface structure of hydrogen-terminated $^{28}$Si(100)--2$\times$1.
	Rows of Si-Si dimers are evident on the surface with each surface Si atom capped by a hydrogen (H) atom.
	The inset depicts a top-down view of the surface and shows the surface's bar-like feature.
	The bars correspond to the Si-Si dimers,
	and rows of Si-Si dimers are separated by gullies comprising Si atoms located one layer below the surface.
	The notation $R_i,i\in\{\text{Si-H, Si-Si, row, dimer}\}$ 
	represents distances between each Si and its neighbors on the surface.
	}
\label{fig:si crystal}
\end{figure}

Each Si atom on a H--$^{28}$Si(100)--2$\times$1 surface shares two bonds with bulk Si, has one dimer bond with another surface Si
and a fourth bond with a surface~H, which can be removed thereby creating a DB.\cite{TW06, ThB95, BB00}
Dangling bonds are readily created on the surface of silicon by using 
a scanning tunneling microscope (STM) to remove hydrogen atoms.
We focus on the well studied case of DBs on the H--$^{28}$Si(100)--2$\times$1 surface~\cite{HPD+09}
depicted in Fig.~\ref{fig:si crystal}.
\emph{Ab initio} calculations and experimental observations reveal that the Si-Si dimer bond is 
$\text{R}_{\text{Si-Si}}\sim2.25$~\AA,\ the Si-H bond is $\text{R}_{\text{Si-H}}\sim1.509$~\AA, 
the distance between adjacent dimers is $\text{R}_{\text{dimer}}\sim3.84$~\AA, 
and the distance between adjacent rows is $\text{R}_{\text{Row}}\sim7.68$~\AA,
which is $\sim\sqrt{2}\times5.431$~\AA\
with $5.431$~\AA\  the unit cell length in Si.\cite{PLH+11}

The DB can be created on a H--$^{28}$Si(100)--2$\times$1 surface by selectively removing 
a surface H atom by applying a voltage pulse using an STM tip.\cite{HPD+09,PLH+11, LXS+10}
Structural and electronic properties of DBs have been studied extensively.\cite{TW06, McEAB98, BB00}

\emph{Ab initio} calculations show that a neutral DB located on a H--$^{28}$Si(100)--2$\times$1 
surface has only one confined electron with an energy located within the Si bandgap.\cite{LXS+10}
This eigenstate is decoupled from the Si conduction and valence bands, 
resulting in a sharply localized electron wavefunction, which makes the DB amenable 
to electronics applications such as a quantum dot.

A DB can lose its single electron to the bulk, thereby becoming a positively 
charged DB$^+$, or it can host one excess electron (with opposite spin) from the bulk, 
hence becoming negatively charged DB$^-$. Losing or acquiring charge depends 
on the type and amount of doping in the host crystal, and on temperature. 
In our case, the crystal has a high concentration of n-type phosphorous (P) doping, 
$\sim5\times10^{18}$~cm$^{-3}$;
therefore,
each DB is highly likely to carry an excess electron.
Due to the on-site Coulombic energy,
the $\text{DB}^{-}$ level is shifted upward in energy by $0.5$ eV
relative to the neutral DB level, but remains within the bandgap.\cite{HPD+09} 

\subsection{Dangling bond pair with a shared excess electron}
\label{subsec:danglingbondpair}

As mentioned above, although an individual isolated DB is highly likely to be negatively charged, 
for DB pairs with a separation  less than 16~\AA, strong Coulombic repulsion between 
the excess electrons on the two DBs destabilizes such a charge configuration (with $2e$ per DB pair). 
The DB pair re-stabilizes by losing one of the excess electrons into the Si conduction band.
The resulting configuration is the DBP$^-$ with the excess electron shared between the two DBs by tunnel coupling.

\emph{Ab initio} calculations show that the two lowest lying energy levels of the DBP$^-$ 
are situated within the silicon bandgap. \cite{LXS+10}
The excess electron tunnels between 
the two DBs with a tunneling frequency $\Delta$, which is a function of the DB separation 
distance as well as of the potential barrier height between them. 
For small DB separations, this barrier is narrow and low, 
only a few tenths of an eV.

In addition to the upper bound for the DBP$^-$ separation, the geometry of 
the surface reconstruction also sets a lower bound,\cite{TW06, HPD+09, HHO+99} see
Fig.~\ref{fig:si crystal}. 
This minimum separation is $\sim 3.84$~\AA, i.e. the dimer-dimer separation along the dimer row. 
These minimum and maximum separations result in an upper and lower bound, respectively,
on the tunneling frequency of the excess electron given by\cite{LXS+10}
\begin{equation}
	\Delta_{\rm max}\sim467~\text{THz}\;,
	\Delta_{\rm min}\sim 0.1~\text{THz}.
\end{equation}

\subsection{Coupled dangling bond pair as a  quantum-dot charge qubit}

A single bound electron shared between left and right DBs can behave as a two-level system,
in the sense that, in the position representation, the states of the system can be given in terms of 
the orthogonal left and right states, or superpositions thereof.
This two-level DBP$^-$ system can be also thought of as a double-well potential
where the individual wells corresponds to left and right DB with quantum states 
$\vert \text{L}\rangle$ and $\vert \text{R}\rangle$, respectively.\cite{HFC+03, GHW05}
This two-level approximation holds if energy levels of each potential well
are widely spaced so that only the ground states of each well are involved in the quantum superposition.

A useful alternative to the left-and-right basis is the basis corresponding to symmetric
and antisymmetric states, $|\psi_+\rangle$ and antisymmetric  $|\psi_-\rangle$, respectively,
which are eigenstates of the Hamiltonian in the absence of biasing fields.
The left and right states can be recovered according to
\begin{equation}
	\vert \text{L}\rangle
		=\frac{1}{\sqrt{2}}\left(\vert\psi_+\rangle-\vert\psi_-\rangle\right),\;
	\vert \text{R}\rangle
		=\frac{1}{\sqrt{2}}\left(\vert\psi_+\rangle+\vert\psi_-\rangle\right).
\end{equation}
Based on \emph{ab initio} calculations, the energies of the symmetric 
and antisymmetric states are within the silicon band gap.\cite{LXS+10}
The higher energy levels are all above the silicon band gap so that an electron 
will become delocalized in the silicon conduction band, rather than assuming a localized
excited state; hence the two-level system approximation, including loss, yields an excellent model.

We are now ready to construct the Hamiltonian for coherent dynamics of the 
charge qubit.
For~$E_{\text{0,L}}$ and~$E_{\text{0,R}}$ the on-site energies plus local 
field corrections for the left and right DBs respectively, the free Hamiltonian for uncoupled DBs is
$E_{\text{0,L}} \vert \text{L}\rangle\langle \text{L}\vert+E_{\text{0,R}}\vert \text{R}\rangle\langle \text{R}\vert$.
We assume that $E_0\equiv E_{\text{0,L}}=E_{\text{0,R}}$ (a symmetry condition)
The Hamiltonian for coherently coupled DBs is
\begin{equation}
\label{eq:Hbare}
	\hat{H}_0
		=E_{\text{0}}\left( \vert \text{L}\rangle\langle \text{L}\vert
			+\vert \text{R}\rangle\langle \text{R}\vert\right )
			+\frac{\hbar\Delta}{2}\left ( \vert \text{R}\rangle\langle \text{L}\vert+\vert \text{L}\rangle\langle \text{R}\vert\right).
\end{equation}
Diagonalizing $\hat{H}_0$ yields eigenenergies~$E_{\text{0}}\pm\hbar\Delta/2$ with corresponding
eigenstates $|\psi_\pm\rangle$, respectively.

\section{Dangling bond pair under static bias and driving laser field}
\label{subsec:appliedDC}

\subsection{Static bias}

Applying a static bias~$V_{\text{b}}$ to DBP$^-$ enables one to control 
the tunneling rate in the system by creating an energy offset 
$e V_{\text{b}}=E_{\text{0,L}}-E_{\text{0,R}}$
between the left and right DB
while preserving the local confinement potential characteristics at each well
as depicted in Fig.~\ref{fig:N1scheme}.
This bias is physically implemented by local electrodes in the vicinity of the DBP$^-$.

\begin{figure}
	\includegraphics[width=\columnwidth]{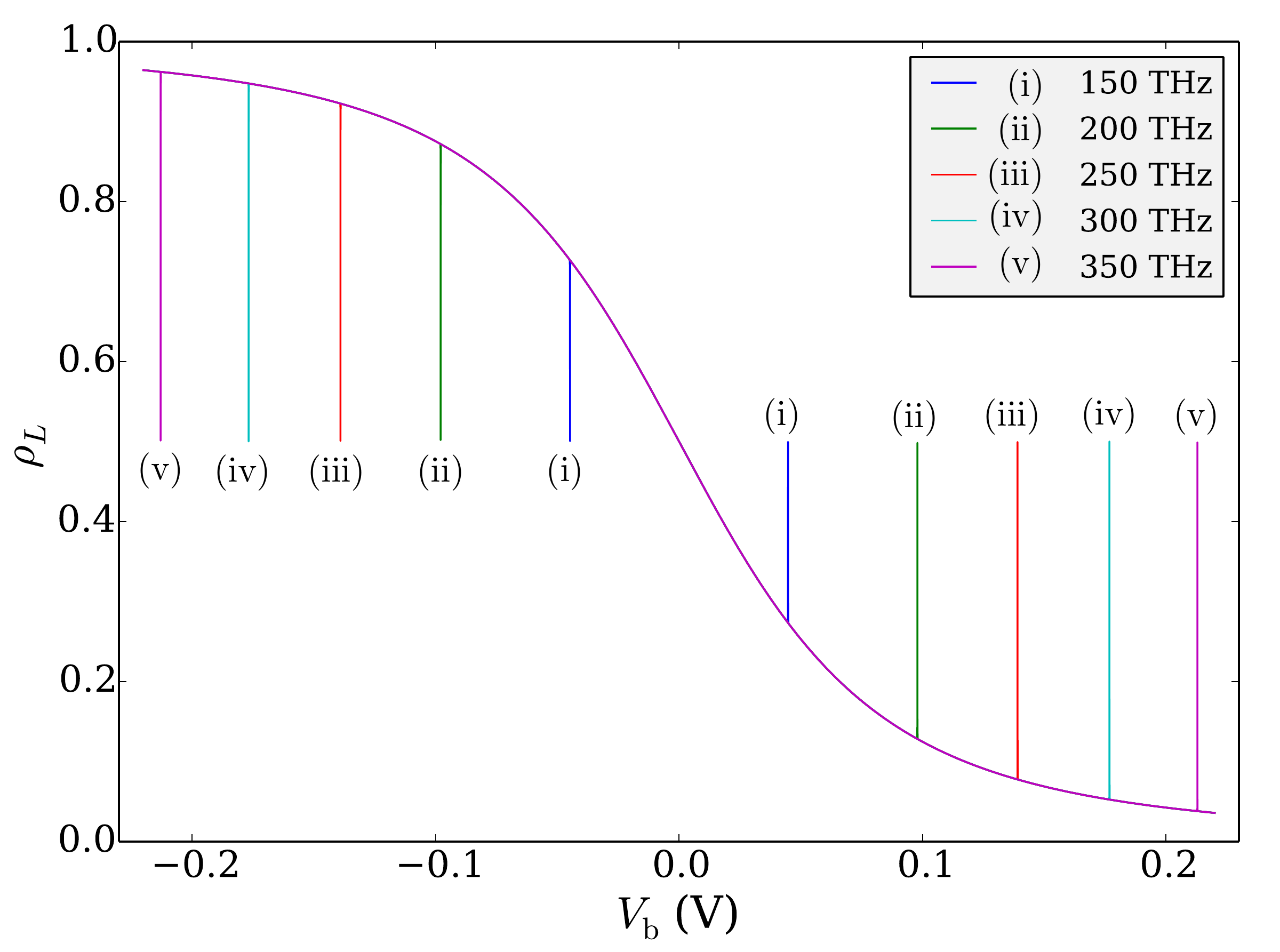}
\caption{
	Time-averaged charge probability in the left DB 
        as a function of applied static bias 
	and under MIR lasers of different frequencies shown in the legend.
	Here, we have chosen $\Delta=133$~THz and
	$\Omega_{\rm MIR}=$ 1 GHz. In the absence of a driving radiation, 
	the result becomes the smooth curve joining all the shown peaks.
	}
\label{fig:rho_vs_bias_with_resonances}
\end{figure}

The Hamiltonian of a DBP$^-$ subjected to a static bias is given by\cite{vdWdFE+02}
\begin{align}
\label{eq:Hsystem}
	\hat H_{\text{b}}=&E_{\text{0}}'\openone
		+\frac{eV_\text{b}}{2}\left(\vert L\rangle\langle L\vert
			-\vert R\rangle\langle R\vert\right)
				\nonumber\\
		&+\frac{\hbar\Delta}{2}\left(|L\rangle\langle R|+|R\rangle\langle L|\right)
\end{align}
with
\begin{equation}
	E_{\text{0}}':=\frac{E_{\text{0,L}}+E_{\text{0,R}}}{2},\;
	eV_\text{b}:=E_\text{0,L}-E_\text{0,R},
\end{equation}
and where $\openone$ is the identity matrix.
Diagonalizing the biased Hamiltonian for the DBP$^-$~(\ref{eq:Hsystem})
yields 

\begin{equation}
\label{eq:Hdiagonalized}
	\hat H_{\text{b}}
		=E_{\text{0}}'\openone+\frac{\hbar\Delta'}{2}\left(|\text{g}\rangle\langle\text{g}|-|\text{e}\rangle\langle\text{e}|\right)
\end{equation}
and 
 \begin{equation*}
 	\vert \text{g}\rangle=\cos\frac{\theta}{2}\vert \text{L}\rangle+\sin\frac{\theta}{2}\vert \text{R}\rangle,\;
	\vert \text{e}\rangle=\cos\frac{\theta}{2}\vert \text{R}\rangle-\sin\frac{\theta}{2}\vert \text{L}\rangle
\end{equation*}
where
 $$\theta=\tan^{-1}\left(\frac{\hbar\Delta}{e V_{\text{b}}}\right).$$
The resultant modified tunnelling frequency is thus
\begin{equation}
\label{eq:modified_delta}
	\Delta'=\sqrt{\Delta^2+\left(\frac{eV_{\text{b}}}{\hbar}\right)^2},
\end{equation}
and the time-averaged charge distribution on the left dot is
\begin{equation}
\label{eq:rho vs bias}
	\rho_{\text{L}}
		=\vert\langle\text{L}\vert\text{g}\rangle\vert^2
		=\cos^2\frac{\theta}{2}
		=\frac{\Delta^2}{\Delta^2+\left(\Delta'+\frac{eV_{\text{b}}}{\hbar}\right)^2},
\end{equation}
which is depicted as the solid black line in Fig.~\ref{fig:rho_vs_bias_with_resonances}. 
As expected, the charge distribution is equal for the two DBs when the 
electric potential bias is zero.

\subsection{Mid-infrared driving field}

We now add a MIR driving field with frequency $\omega_{\text{MIR}}$ to the scheme 
as shown in Fig.~\ref{fig:N1scheme}.
The purpose of this driving field is to probe the system to discover resonance
conditions whereby the barrier between the left and right DBs effectively 
vanishes.\cite{PJM+04} A tunable continuous-wave 
solid-state laser or CO$_2$ gas laser are examples of suitable MIR sources.\cite{CT02} 
The MIR beam intensity must be weak 
enough to ensure that multi-photon resonances are negligible,\cite{PJM+04, SDB05}
but strong enough to drive the oscillation between left and right DBs.

A quantitative description of the dynamics begins by treating the biased DBP$^-$ 
as an electric dipole with an approximate transition dipole moment 
$\bm{d}_{\text{DBP}^-}=-e\bm{x}$, which is the 
product of the electron charge $e$  and the inter-DB separation vector $\bm{x}$
pointing from the negative to the neutral DB. The corresponding dipole-moment operator is 
$\hat{\bm{d}}=\bm{d}_{\text{DBP$^-$}} \hat\sigma_x$.

The electric-dipole interaction with the MIR electric field
$\bm{E}_{\text{MIR}}$ is given by the interaction Hamiltonian~\cite{AE75}
\begin{equation}
	\hat{H}_{\text{dipole}}=-\hat{\bm{d}}\cdot\bm{E}_{\text{MIR}}.
\end{equation}
The interaction strength is quantified by the Rabi frequency
\begin{equation}
\label{eq: Rabi frequency}
	\Omega_{\text{MIR}}=\frac{\hat{\bm{d}}\cdot\bm{E}_{\text{MIR}}}{\hbar},
\end{equation}
and the resultant driving-field interaction Hamiltonian is~\cite{BS06}
\begin{align}
\label{Hdrive}
	\hat H_\text{d}
		=&\hbar\left|\Omega_\text{MIR}\right|\cos\omega_\text{MIR}t
			\big[\cos\delta\left(|\text{R}\rangle\langle\text{R}|
				-|\text{L}\rangle\langle\text{L}|\right)
					\nonumber\\
			&+\sin\delta\left(|\text{L}\rangle\langle\text{R}|
				+|\text{R}\rangle\langle\text{L}|\right)\big]
\end{align}
for~$\delta$ a parameter containing information about the laser beam angle and ratio 
of wavelength to dipole length.

The intensity of the radiation is related to the Rabi frequency via  the electric-field amplitude,
$I=\epsilon_0\vert\bm{E}\vert^2 c$
with $\epsilon_0$ the permittivity,
$\bm{E}$ the electric-field amplitude, 
and~$c$ the speed of light.
Assuming that $\bm{E}_\text{MIR}$ and~$\bm{d}_{\text{DBP}^-}$ are parallel,
Eq.~(\ref{eq: Rabi frequency}) yields
\begin{equation}
	|\bm{E}|=\hbar\Omega_\text{MIR}/|\bm{d}_{\text{DBP}^-}|.
\end{equation}
For a DBP$^-$ with inter-DB distance
$\vert\bm{x}\vert= 7.68$ \AA, we obtain
$\bm{d}_{\text{DBP}^-}\approx 10^{-28}$ Cm.

\begin{figure}
	\includegraphics[width=\columnwidth]{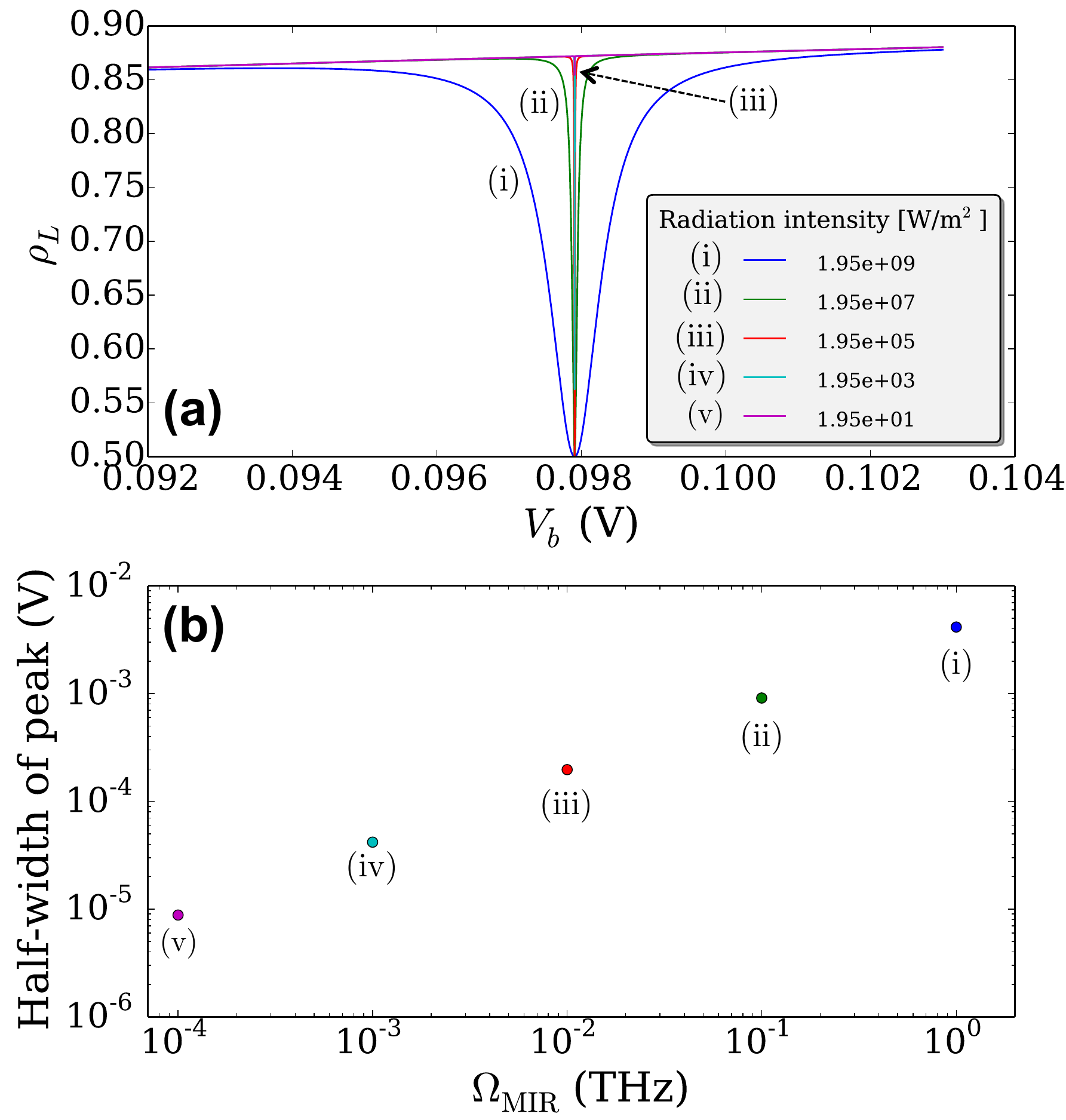}
	\caption{
	The effects of varying the driving field intensity on the resonance peak shapes.
	(a)~Here $\omega_{\rm {MIR}}$ is kept fixed at 200 THz, and as the field intensity is reduced the
resonance peaks become narrower. (b) The peak widths are extracted from the above plots
and are plotted against the corresponding Rabi frequencies. 
The color of each point here corresponds
to a plot in~(a).
The bottom-left point corresponds 
to the lowest measurable field intensity as limited by the thermal noise in the biasing electrodes.}
\label{fig:laser_intensity_effect}
\end{figure}

In addition to the conditions above,
the Rabi frequency is low enough to avoid multi-photon resonances
but high enough to drive the oscillation, and the choice
$\Omega_\text{MIR}/\omega_{\text{MIR}}$ offers some flexibility to tune these parameters.
We therefore studied the effects of the intensity of the applied laser in 
Fig~\ref{fig:laser_intensity_effect}, where we varied this quantity over
a few orders of magnitude.

We can see in this figure that, as the intensity is decreased, the width
of the resonance peaks decreases as well. Eventually the width
becomes lower than the noise in the applied bias, for a laser intensity 
of about 20 W/m$^2$.
The horizontal resolution of the resonance was estimated
by treating the width as being due to thermal noise in the biasing electrodes,
namely the Johnson-Nyquist noise given by the formula
\begin{equation}
	V_\text{JN}= \sqrt{4kTRB}
\end{equation}
for $R = 1$~M$\Omega$,
and $B = 3$~kHz, similar to values present in an STM instrumental setup.
The corresponding Rabi frequencies for each peak width
are plotted in Fig~\ref{fig:laser_intensity_effect}(b) (anticipating that this is also the range of useful 
Rabi frequencies for the purpose of this study, i.e.\ 0.1 GHz to 1 THz.)

Corresponding radiation intensities ranging from 20 to $10^{9}$ W/m$^2$) are experimentally 
feasible with current CO$_\text{2}$ lasers.
CO$_\text{2}$ lasers are high-power, continuous-wave lasers that generate 
infrared light with wavelength~$\lambda$ within the domain of 9.2--11.4 $\mu$m,
corresponding to a frequency range of 165--205 THz, and have an operating 
power~$P_{\text{CO}_\text{2}}$ from mW to hundreds of W.

\subsection{The combined action of static bias and driving field}

The Hamiltonian including both the static bias and the driving field is
\begin{equation}
	\hat H_\text{bd}=\hat H_{\text{b}}+\hat H_{\text{d}}.
\end{equation}
Converting to the interaction picture according to
\begin{equation*}
	\hat{H}_\text{I}
		=U^\dagger \hat{H}_\text{bd} U,\;
	U:=\exp\left[-\text{i}\frac{\omega_\text{MIR}t}{2}
	\left(|\text{R}\rangle\langle\text{R}|-|\text{L}\rangle\langle\text{L}|\right)\right]
\end{equation*}
eliminates explicit time dependence.
If the detuning $\eta:= \omega_{\text{MIR}}-\Delta'$
is small compared to the frequency sum $\omega_{\text{MIR}}+\Delta'$,
then~\cite{BS06}
\begin{equation}
\label{eq:InteractionHamiltonian}
	\frac{\hat H_{\text{I}}}{\hbar}
		\approx\frac{\Omega}{2} \left(|\text{g}\rangle\langle\text{e}|+|\text{e}\rangle\langle\text{g}|\right)
			+\frac{\eta}{2} \left(|\text{g}\rangle\langle\text{g}|-|\text{e}\rangle\langle\text{e}|\right),
\end{equation}
where
\begin{equation}
	\Omega:=\left|\Omega_{\text{MIR}}\right|\sin\left(\theta-\delta\right)
\end{equation}
for a weak MIR field, namely $\Omega_{\text{MIR}}\ll \omega_{\text{MIR}}$.\cite{AE75}

The eigenenergies of the approximate interaction Hamiltonian~$\hat H_{\text{I}}$ 
in~(\ref{eq:InteractionHamiltonian}) are~\cite{BS06}
\begin{equation}
\label{eq:H_Ieigenenergies}
	\varpi_\pm :=\pm\frac{\hbar}{2}\sqrt{\Omega^2+\eta^2},
\end{equation}
which represent the modified Rabi frequencies.
The corresponding eigenstates are
 \begin{align}
 	\vert \text{g}\rangle_\varphi=&\cos\frac{\varphi}{2}\vert \text{g}\rangle-\sin\frac{\varphi}{2}\vert \text{e}\rangle,
		\nonumber\\
	\vert \text{e}\rangle_\varphi=&\cos\frac{\varphi}{2}\vert \text{e}\rangle+\sin\frac{\varphi}{2}\vert \text{g}\rangle
\end{align}
for
\begin{equation}
	\varphi=\tan^{-1}\left(\frac{|\Omega|}{\eta}\right).
\end{equation}
The probability for the charge to be on the left DB is
\begin{equation}
\label{eq:rho vs bias vs driven}
	\rho_{\text{L}}
		=\vert\langle\text{L}\vert\text{g}\rangle_\varphi\vert^2
		=\frac{1}{2}
			+ \frac{eV_{\text{b}}}{2\hbar\Delta'}
			\frac{1}{1+\frac{1}{2}\tan^2\varphi}
\end{equation}
analogous to the undriven distribution~(\ref{eq:rho vs bias}).

In an actual experiment, for a given MIR frequency~$\omega_\text{MIR}$, the potential bias~$V_\text{b}$
is adjusted until a resonance is found whereby the charge distribution is equal on the two DBs.
Thus, one is expected to obtain curves similar to those in  Fig.~\ref{fig:rho_vs_bias_with_resonances}, 
where the spikes correspond to cases that $\omega_{\text{MIR}}=\Delta'$.
Intuitively, these resonances correspond to the MIR driving field overwhelming the biasing field,
effectively making the barrier negligible and the charge distribution equal in either DB.

\begin{figure}
	\includegraphics[width=\columnwidth]{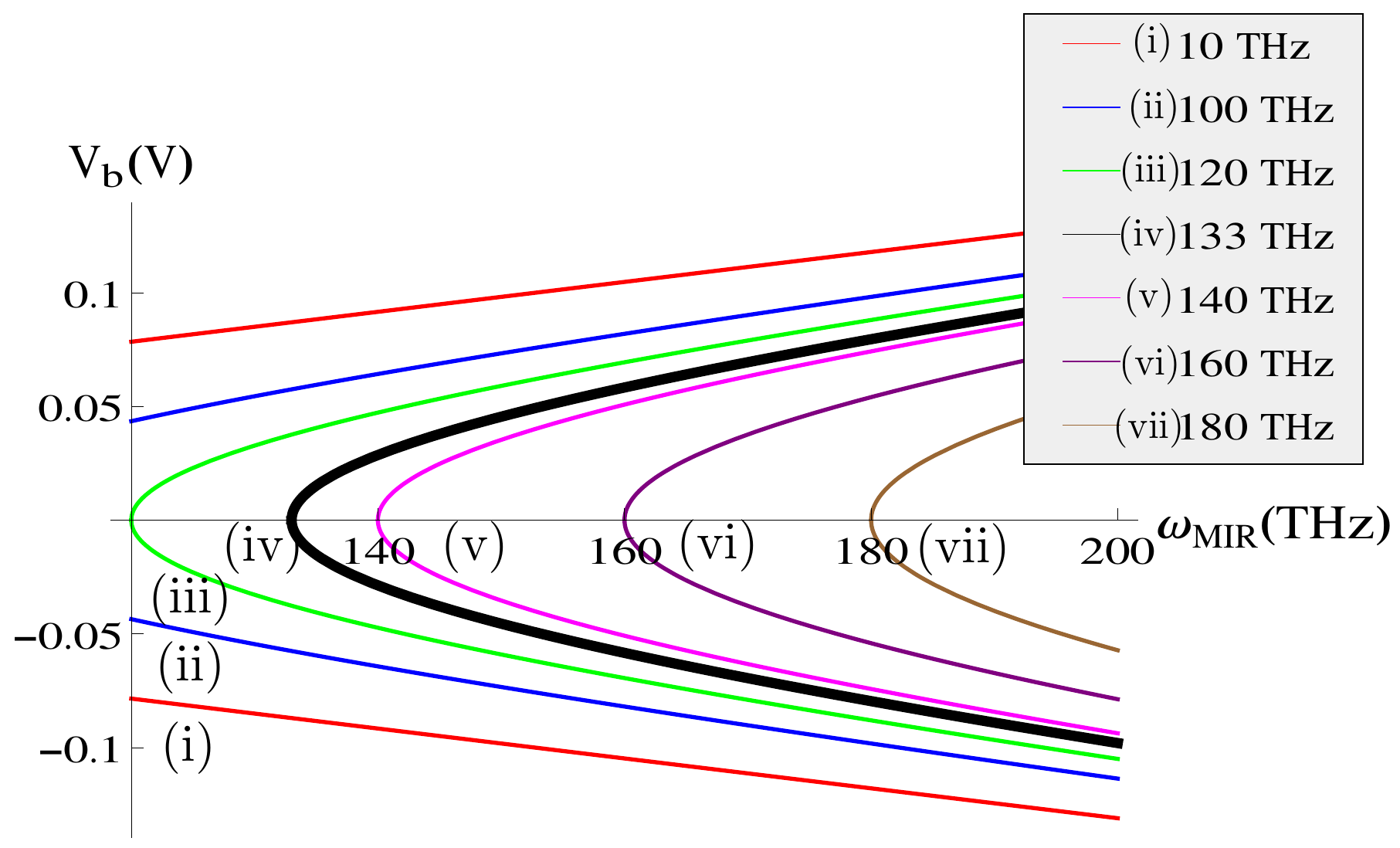}
\caption{
	Contour plots  of loci in the parameter space ($V_\text{b}$, 
	$\omega_\text{MIR}$) where resonances occur, 
	i.e. spikes in Fig.~\ref{fig:rho_vs_bias_with_resonances} where $\rho_\text{L}=\frac{1}{2}$. 
	Each contour corresponds to a different native DBP$^-$ tunneling frequency~$\Delta$, indicated in the legend.
	}
\label{fig:2Dbias_vs_omegaNIR}
\end{figure}

Figure~\ref{fig:2Dbias_vs_omegaNIR} shows an extended parameter space as contour plots 
of loci where, for fixed values of~$\Delta$ indicated in the legend, resonances such as 
those in Fig.~\ref{fig:rho_vs_bias_with_resonances} occur.
This figure illustrates a key point of our scheme, namely that the MIR frequency 
and potential bias can be tuned to discover the tunneling frequency~$\Delta$ 
simply by measuring the probability of the excess charge being in the left DB.

\section{Mitigating laser heating}
Excessive heating by the incident laser radiation can lead to
damage of the sample.
An important detail when estimating laser damage in our sample is that we only require 
radiation with sub-bandgap energy, for which the silicon absorption coefficient is relatively small. 
Therefore, silicon crystals are resilient to heating and have a high thermal and optical damage threshold\cite{RSS+06} 
in the MIR range of interest due to low absorption for this spectral domain. 

For the H--Si(100) substrate in our study, above-bandgap driving radiation 
($\lambda$= 532 nm) with an intensity of 2.67$\times 10^{11}$ W/m$^2$ suffices
to cause hydrogen desorption but does not damage the sample.\cite{schwalb2007real}
Given that we require sub-bandgap radiation ($\lambda$= 9-19 $\mu$m) with two orders of magnitude lower intensity,  any damage to the sample is expected to be highly unlikely.
At these greatly reduced intensity levels, hydrogen desorption is also unlikely.~\cite{schwalb2007real}

However, for a real sample the specific type and number of defects might be
generally unknown, and other heating mechanisms might be at play.
For the case when a high Rabi frequency (1 THz and higher) is required
in order to induce measurable resonance peaks, the correspondingly
high laser intensity might be of practical concern inasmuch
as a prolonged laser exposure is required during the experiment.
Therefore, we here estimate theoretically the effects of laser heating under
those conditions, and suggest ways to deal with them.

The flow of heat can be modeled by the heat
equation~\cite{CJ59}
\begin{equation}
\rho c_{p}\frac{\partial T\left(\bm{r},t\right)}{\partial t}-\kappa\nabla^{2}T\left(\bm{r},t\right)=Q_\text{L}\left(\bm{r}\right)
\end{equation}
where $\bm{r}=\left(x,z\right)$, $T\left(\bm{r},t\right)$ is the temperature function, $\rho$ is the
mass density, $c_{p}$ is the specific heat, $\kappa$ is the thermal conductivity, and 
$Q_\text{L}\left(\bm{r}\right)$ is the heat-source term related to the 
power density injected by the laser into
the sample.
Note that the material constants appearing in these coefficients
are generally functions of temperature, but here they were assumed to be constant.

The source term varies spatially with the depth into the sample, according
to the power absorbed from the incident laser as
\begin{equation}
	Q_\text{L}\left(\bm{r}\right)
		=Q_\text{L}\left(z\right)
		=\alpha\tau I_{0}\text{e}^{-\alpha \left(z-z_0\right)}
\end{equation}
where $\alpha$ is the decay rate of the radiation intensity in the
sample, $\tau$ is the transmission coefficient at the surface,
$I_{0}$ is the intensity of the incident radiation, and $z_0$ is the location of the surface. 

We solve the boundary value problem (BVP) consisting of the above heat equation
and appropriate boundary conditions (BCs) by the finite element method
on a two-dimensional $\left(x,z\right)$-grid spanning a $10\times10~\mu\text{m}$ region.
We solve this BVP for consecutive time slices using an Euler scheme
for sampling time evolution. We use the esys-escript FEM library~\cite{GBH+}
for getting the numerical solution.

\begin{figure}
	\includegraphics[width=\columnwidth]{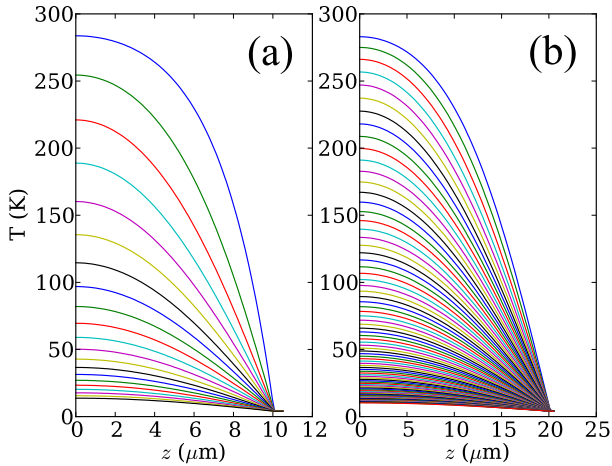}
\caption{
	Effects of laser heating: temperature profiles ($T$) as functions of depth ($z$)
	into the sample for different times from the moment the laser was
	turned on. 
	Curves show increasing time slices from top to bottom in increments of 0.1 $\mu$s. 
	The initial temperature is 300~K at the surface and 4~K at the back plane.
	(a) The cool boundary is placed at 10~$\mu$m inside the sample and
	the bottommost temperature profile is reached at 2$\mu$s. 
	(b) The cool boundary is placed at 20~$\mu$m inside; 
	the bottommost temperature profile is reached at 9$\mu$s.
	}
\label{fig:Effects-of-laser}
\end{figure}

Given that the choice of BCs has a drastic effect on the
evolution of the temperature in our system, we explored various setups
for the purpose of minimizing the heating effect. We find the most
favorable BC to be a Dirichlet-type condition at the back of the
sample, where we set the temperature to a fixed value of 4 K, i.e.
simulating the effect of a liquid-helium-cooled metal plate in thermal
contact with the back of the Si sample. For this setup, we obtain
the time evolution shown in Fig.~\ref{fig:Effects-of-laser} for the temperature
profile as a function of the depth into the sample.

A laser of total power 1 W is focused on an area of $10\times10$ $\mu$m$^{2}$,
corresponding to an intensity of $10^{10}$~W/m$^2$, chosen higher than any 
intensity expected in the actual experimental situation.
Material parameters
used in the simulation are the following: thermal conductivity of Si: 149 W/K-m;
mass density of Si: 2329 kg/$\mathrm{m}{}^{2}$; specific heat of
Si: 814 J/kg-K; complex refractive index of Si: $n+ik$ with $n$
= 3.4215, $k=6.76\times10^{-5}$ for a frequency of 100 THz.\cite{Palik1985}
Note that the above value of the refractive index corresponds to low-
to medium-doped Si.
In general, the value of the absorption coefficient 
in the infrared range
depends strongly on the doping level and can become greater than 
2000 cm$^{-1}$ for donor concentrations above 10$^{19}$ cm$^{-3}$.\cite{SF57}
However, in our proposed setup, doping is only practically required in a very thin topmost layer, 
resulting in negligible heating, as calculated.

In order to eliminate any doubts about the role of the exact thermal resistivity
gradient at the cooler boundary, we simulated the same system with 
this boundary placed at different depths from the silicon surface, 10 and 20 $\mu$m, 
with results shown in Fig.~\ref{fig:Effects-of-laser} (a) and (b), respectively.
As expected, the thicker sample in (b) takes a few times longer to cool, but it
ends up saturating to a cool profile nonetheless. 
The same cooling behavior is present for a cool reservoir at 4~K or 77~K.

To sum up, our simulations show that even in the presence of laser radiation,
our sample cools off to a steady profile in a few microseconds
if the cool boundary is placed a few tens of $\mu$m from the silicon surface. 
Therefore we expect excessive heating can be avoided by using a cooling system
even in cases with the most intense radiation required for our experimental apparatus.

\section{Atomic force microscopy characterization of tunneling between dangling
bonds}\label{sec:AFM}

AFM is ideally suited to measure the spatial charge
distribution in a DBP$^{-}$ (Fig.~\ref{fig:N1scheme})
without significantly distorting the electronic landscape of the sample. 
The AFM has been shown to detect single charges.\cite{CMG09,SMS+05}

\subsection{Modeling atomic force microscope in frequency modulation mode}

\label{subsec:modeling}
The AFM cantilever behaves as a simple harmonic oscillator along the coordinate
axis $z$ perpendicular to the sample surface. The tip is driven
by an externally controlled force $F_{0}\sin\omega_{0}t$, with $F_{0}$
being constant. When scanning a sample, the AFM tip experiences distance-dependent
forces $F_{\text{z}}\left(z\right)$ from its interaction with the sample.

In the limit of small oscillation amplitudes and small force gradients,
the equation of motion for the AFM tip around its equilibrium position
(chosen as the origin of the $z$-axis, at a height~$z_{0}$ from
the surface) is~\cite{MDS+08} 
\begin{equation}
m\ddot{z}+\gamma\dot{z}+m\omega_{0}^{2}z=F_{0}\sin\omega_{0}t+F_{\text{z}}\left(z\right),\label{eq:unharmoniceq}
\end{equation}
where $\gamma$ is the AFM damping factor, $m$ the mass of the probe
and
\begin{equation}
	k=m\omega_{0}^{2}
\end{equation}
the AFM probe spring constant. 

In the same limit of small oscillation amplitudes (a few~\AA~ is anticipated), we can use a truncated
Taylor expansion 
\begin{equation}
F_{\text{z}}\left(z\right)\simeq F_{\text{z}}\left(0\right)+z\left.\frac{\partial F_{\text{z}}}{\partial z}\right\vert _{z=0},\label{eq:taylor}
\end{equation}
with a resultant equation of motion for the tip, 
\begin{equation}
m\ddot{z}+\gamma\dot{z}+m\omega^{2}z\simeq F_{0}\sin\omega_{0}t+F_{\text{z}}\left(0\right),\label{eq:AFMtipSHMmodifiedfreq}
\end{equation}
describing driven oscillations with a modified resonant frequency
depending on the lateral tip position 
\begin{equation}
\omega^{2}=\omega_{0}^{2}-\frac{1}{m}\left.\frac{\partial F_{\text{z}}}{\partial z}\right\vert _{z=0}.\label{eq:modifiedk}
\end{equation}
The right-hand side of Eq.~(\ref{eq:AFMtipSHMmodifiedfreq}) is a
constant in space so the tip-sample force is detected by measuring the
frequency response of the tip according to~(\ref{eq:modifiedk}).
Employing the binomial expansion on Eq.~(\ref{eq:modifiedk}) yields
the modified frequency expression 
\begin{equation}
\Delta\omega:=\omega-\omega_{0}\simeq-\frac{\omega_{0}}{2k}\left.\frac{\partial F_{\text{z}}}{\partial z}\right\vert _{z_{0}}\label{eq:freqshift}
\end{equation}
showing the proportionality between the frequency shift and the local
force gradient. 

Note that far from the limit of small amplitudes one
can still approximate the AFM motion from the above equation, but instead
of the force gradient at the equilibrium position, one should use an average force
gradient over an entire oscillation range $[z_{\mathrm{min}},z_{\mathrm{max}}]$,
i.e.
\begin{equation}
\Delta\omega\simeq-\frac{\omega_{0}}{2k\left(z_{\mathrm{min}}-z_{\mathrm{max}}\right)}\intop_{z_{\mathrm{min}}}^{z_{\mathrm{max}}}\frac{\partial F_{\text{z}}}{\partial z}\text{d}z.
\end{equation}
The experimental goal is then to measure these changes in the tip
oscillation frequency thereby revealing information about the sample.

From Eq.~(\ref{eq:freqshift}), we see that the ratio $\omega_{0}/k$
gives the sensitivity of the cantilever, which in practice depends on
the build geometry and material of the cantilever. Typical examples are
silicon cantilevers with a sensitivity factor $\omega_{0}/k=4000$ 
Hz m/N, and the qPlus tuning fork with $\omega_{0}/k=20$ Hz m/N.\cite{MGW09,Gie03,GML+09}
However, when choosing a cantilever for a given experiment, the sensitivity
is not the only factor to consider, as scan stability (e.g. against
jump-to-contact), quality factor, measurement bandwidth, and appropriate
size of oscillation amplitudes also play important roles. 

The minimum detectable signal of an AFM experimental setup is determined by assessing its frequency noise $\delta\left(\Delta\omega\right)$, i.e. the standard deviation of the frequency shift. Theoretically,  $\delta\left(\Delta\omega\right)$ is given by\cite{SchWG12, KYM11}
\begin{equation}\label{eq:freq noise}
\delta\left(\Delta\omega\right)=\frac{2\pi}{A}\sqrt{\frac{\omega_{0}Bk_{\mathrm{B}}T}{2\pi^{2}kQ}
+\frac{n_{q}^2B^{3}}{\pi^2}
+\frac{n_{q}^2 B}{2 Q^2}},
\end{equation}
where  $Q$ is the quality factor, $A$ is the oscillation amplitude,
$B$ is the measurement bandwidth, $k_{\mathrm{B}}T$ is the thermal
energy, and $n_{q}$ is the deflection noise density.  
The first term on the right-hand side of Eq.~(\ref{eq:freq noise}) is the thermal noise of the AFM tip, the second term is the deflection-detector noise, and the third term is the noise of the instrumental setup.
Thus, one should choose the experimental parameters such that the sensitivity and the signal-to-noise ratio of the AFM setup are optimized.

\subsection{Tip-sample interactions}\label{subsec:tip-sample force interaction}

In using the AFM to measure the charge distribution at the silicon surface, 
all significant tip-sample forces must be considered by our model. These
can be short-range chemical forces (less than 5~\AA{}), long-range
van der Waals forces, electrostatic forces, or magnetic forces (up
to 100 nm).\cite{Gie03} However, we choose operational parameters 
to ensure that electrostatic forces produced by the DBP$^{-}$ dominate over these other forces.

The external source potential~$V_{\text{t}}$ is kept
constant during the interaction with the sample. If the tip is sufficiently far from the surface, chemical forces can be ignored, and magnetic
forces are negligible if the tip is made of non-magnetic material,
e.g.~tungsten. An ultrasharp nanotip~\cite{RPW06}
is employed to minimize forces arising from induced polarization
of the sample. Reducing the tip oscillation amplitude to the \AA{}ngstrom
scale, for example with a quartz-made qPlus sensor,\cite{MGW09,Gie03,GML+09}
also helps to minimize this form of interaction.

\subsection{Trapped charge in the tip-sample system}

\label{subsec:trapped} 

\subsubsection{Electrostatic potential energy}

\begin{figure}
	\includegraphics[width=\columnwidth]{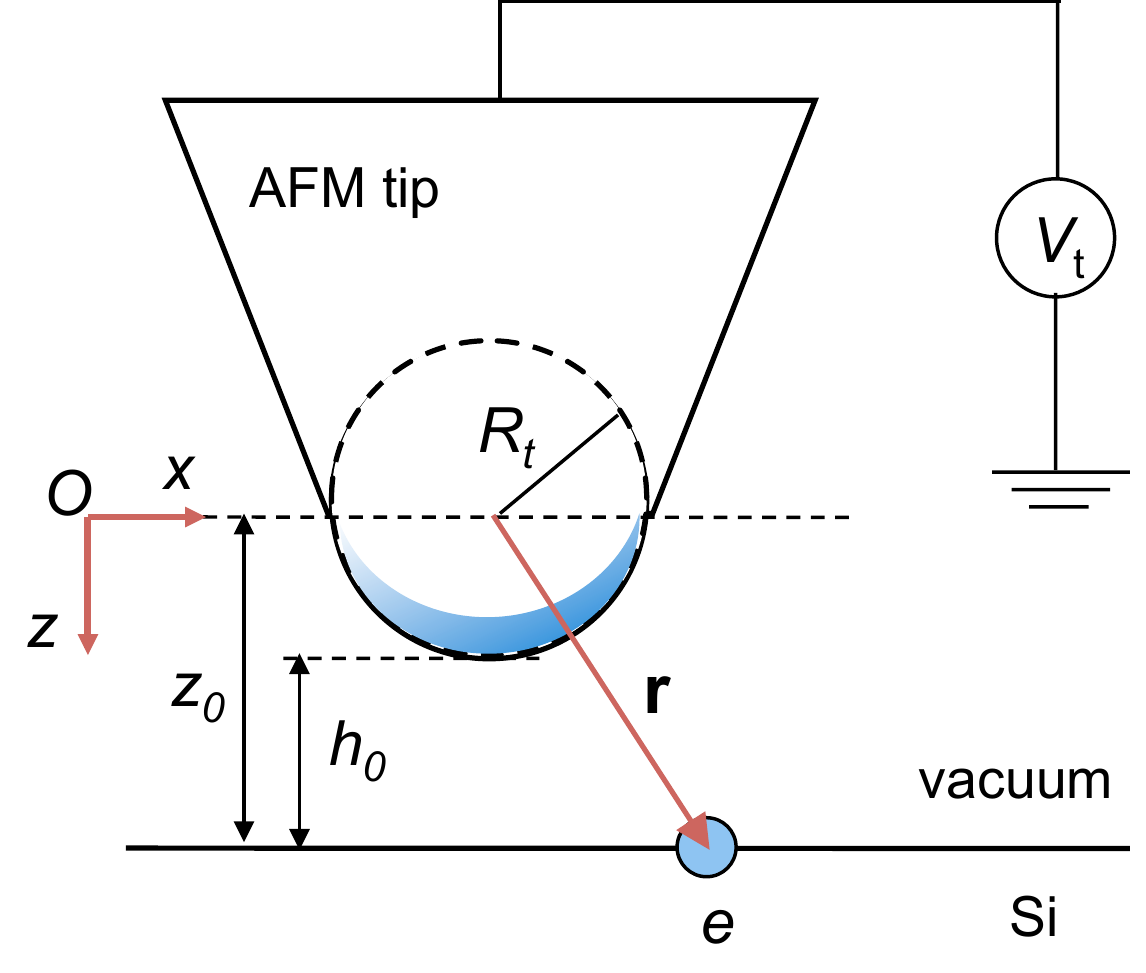}
\caption{Schematic of AFM setup for charge sensing, illustrating 
the geometrical parameters relevant for the tip-sample interactions.
The choice of the coordinate system ($xOz$), the position vector of the DB charge ($\bm{r}$),
and the ``boss sphere'' (dashed circle) fitted to the apex of the AFM tip are shown.}
\label{fig:newfig6}
\end{figure}

The DBP$^{-}$ can be treated as a trapped charge oscillating between
the left and right DBs. This single charge is located within the plane
of the Si surface (Fig.~\ref{fig:newfig6}). Thus, the charge interacts
with other entities such as the space-charge layer in the semiconductor 
substrate and other charges in the substrate and on the biased AFM tip, 
if present. Here we analyze the problem from a
basic viewpoint in order to capture the essential electrostatic
elements at play. 

For an n-type Si sample with a donor concentration$N_{\text{D}}$, the presence
of a locally planar electrode at a height $h_{0}$ above the surface biased
at a voltage $V_\text{t}$ induces a subsurface space-charge layer in the
semiconductor with a width $w$ approximated by the solution
of the quadratic equation\cite{F03} 
\begin{equation}
\frac{eN_{D}}{2\varepsilon_{r}h_{0}}w^{2}+eN_{D}w+\frac{\varepsilon_{0}V_{\text{t}}}{h_{0}}=0,
\end{equation}
where $\varepsilon_{0}$ and $\varepsilon_{r}$ are the vacuum permittivity and semiconductor dielectric constant, respectively.

Correspondingly, the so-called band bending potential at a depth $z_\text{d}$
into the sample can be written in the quadratic approximation as~\cite{F03}
\begin{equation}
V_{\mathrm{Si}}\left(z_\text{d}\right)= V_{0}\left(1-\frac{z_\text{d}}{w}\right)^{2}
\end{equation}
where $V_{\mathrm{0}}$ is the potential at the surface given by
\begin{equation}
V_{\mathrm{0}}\approx\frac{eN_{D}w^{2}}{2\varepsilon_{0}\varepsilon_{r}}\text{sign}\left(V_{\text{t}}\right)
\end{equation}
where `sign' gives the sign of the applied tip bias. 
As the AFM tip is usually not locally planar the above equation is a coarse approximation 
for band bending  representing an upper limit for the real case.

In order to calculate the potential at the Si surface, we
approximate the electrostatic potential due to the biased AFM tip as being
that of a biased conducting sphere with radius $R_\text{t}$ fitted to the apex region of the 
tip (or the ``boss'' as depicted in Fig.~\ref{fig:newfig6}).
In order to reflect the contribution of the mobile charge carriers in the substrate, we apply a
rescaling of this spherical potential, namely we recalibrate the value
of the potential at the Si surface location directly under the tip apex,
$\bm{r}_{0}=\left(0,z_{0}\right)$, to be just $V_{\mathrm{0}}$ given above.

From this analysis, at the location
$\bm{r}=\mathit{\left(x,z\right)}$ of the DB, the \emph{bare} potential
due to the tip is 
\begin{equation}
\phi\left(r\right)=\frac{V_{0}r_{0}}{r}
\end{equation}
where the coordinate origin is chosen at the center of the boss
sphere. This bare
potential does not include the image charge effects, which are accounted
for below. Furthermore, for the case when the amplitudes of the AFM
cantilever are small, we can neglect the variation of $V_{0}$ with
the tip height and use henceforth only its value at the equilibrium
scanning height.

With the above assumptions, the effective electrostatic energy of
the tip-charge system can be written as \cite{KLS00}
\begin{equation}
W^{\mathrm{eff}}\left(r\right)=-\frac{1}{2}C_\text{t}V_{\text{t}}^{2}+\frac{eV_{0}z_{0}}{r}-\frac{1}{8\pi\varepsilon_{0}}\frac{e^{2}R_{\text{t}}}{r^{2}-R_\text{t}^{2}}\label{eq:energy1}
\end{equation}
where the last term accounts for the image charge inside the tip and
for the charge redistribution via the voltage source as explained
by Kantorovich et al.\cite{KLS00} (note the change in the unit system used here). 
Then the force exerted on the tip in the direction normal to the surface
can be calculated as
\begin{equation}
F_{z}\left(r\right)=-\frac{\partial W^{\mathrm{eff}}}{\partial z}=\frac{eV_{0}z_{0}z}{r^{3}}-\frac{1}{4\pi\varepsilon_{0}}\frac{e^{2}R_\text{t}z}{\left(r^{2}-R_\text{t}^{2}\right)^{2}}\cdot\label{eq:force1}
\end{equation}
This expression for force can then be
substituted into Eq.~(24) to approximate the expected AFM frequency
shift.

\subsubsection{Atomic-force-microscope frequency shift}

The AFM frequency shift is
obtained from the derivative of the force with respect to $z$, as
in Eq.~(\ref{eq:freqshift})
\begin{eqnarray}
\Delta\omega & = & \frac{1}{2m\omega_{0}}\left[-\frac{eV_{0}z_{0}\left(x^{2}-2z_{0}^{2}\right)}{r^{5}}\right.\nonumber \\
 &  & \left.+\frac{1}{4\pi\varepsilon_{0}}\frac{e^{2}R_\text{t}\left(R_\text{t}^{2}-x^{2}+3z_{0}^{2}\right)}{\left(R_\text{t}^{2}-r^{2}\right)^{3}}\right].
\end{eqnarray}
%

In readout of a DBP$^{-}$ excess charge, the total AFM frequency
shift is given by 
\begin{align}
\Delta\omega_{\text{AFM}} & =\rho_{\text{L}}\Delta\omega^{\left(\text{L}\right)}+\rho_{\text{R}}\Delta\omega^{\left(\text{R}\right)}\nonumber \\
 & =\rho_{\text{L}}\left(\omega^{\left(\text{L}\right)}-\omega_{0}\right)+\left(1-\rho_{\text{L}}\right)\left(\omega^{\left(\text{R}\right)}-\omega_{0}\right)\label{eq:total freq}\\
 & =\xi\rho_{\text{L}}+\left(\omega^{\left(\text{R}\right)}-\omega_{0}\right),\nonumber
\end{align}
where $\Delta\omega^{\left(\text{L}\right)}$ ($\Delta\omega^{\left(\text{R}\right)}$)
is the frequency shift due to the charge localized in the left (right)
DB and each frequency shift is weighted by the corresponding time-averaged
charge probability $\rho_{\text{L}}$ $\left(\rho_{\text{R}}\right)$. The parameter $\xi$ is the differential frequency
shift of the cantilever caused by the excess charge tunneling from
the right to the left DB; i.e. 
\begin{equation}
\xi=\omega^{\left(\text{L}\right)}-\omega^{\left(\text{R}\right)}.\label{Eq:xi}
\end{equation}
Equation~(\ref{eq:total freq}) indicates that the AFM readout $\Delta\omega_{\text{AFM}}$
is linear in $\rho_{\text{L}}$. Thus, we expect to observe resonances
in the AFM output signal while scanning through a range of bias values~$V_{\text{b}}$,
owing to the existence of resonance features
for~$\rho_{\text{L}}$ as seen in Fig.~\ref{fig:rho_vs_bias_with_resonances}. 

\begin{figure}
	\includegraphics[width=\columnwidth]{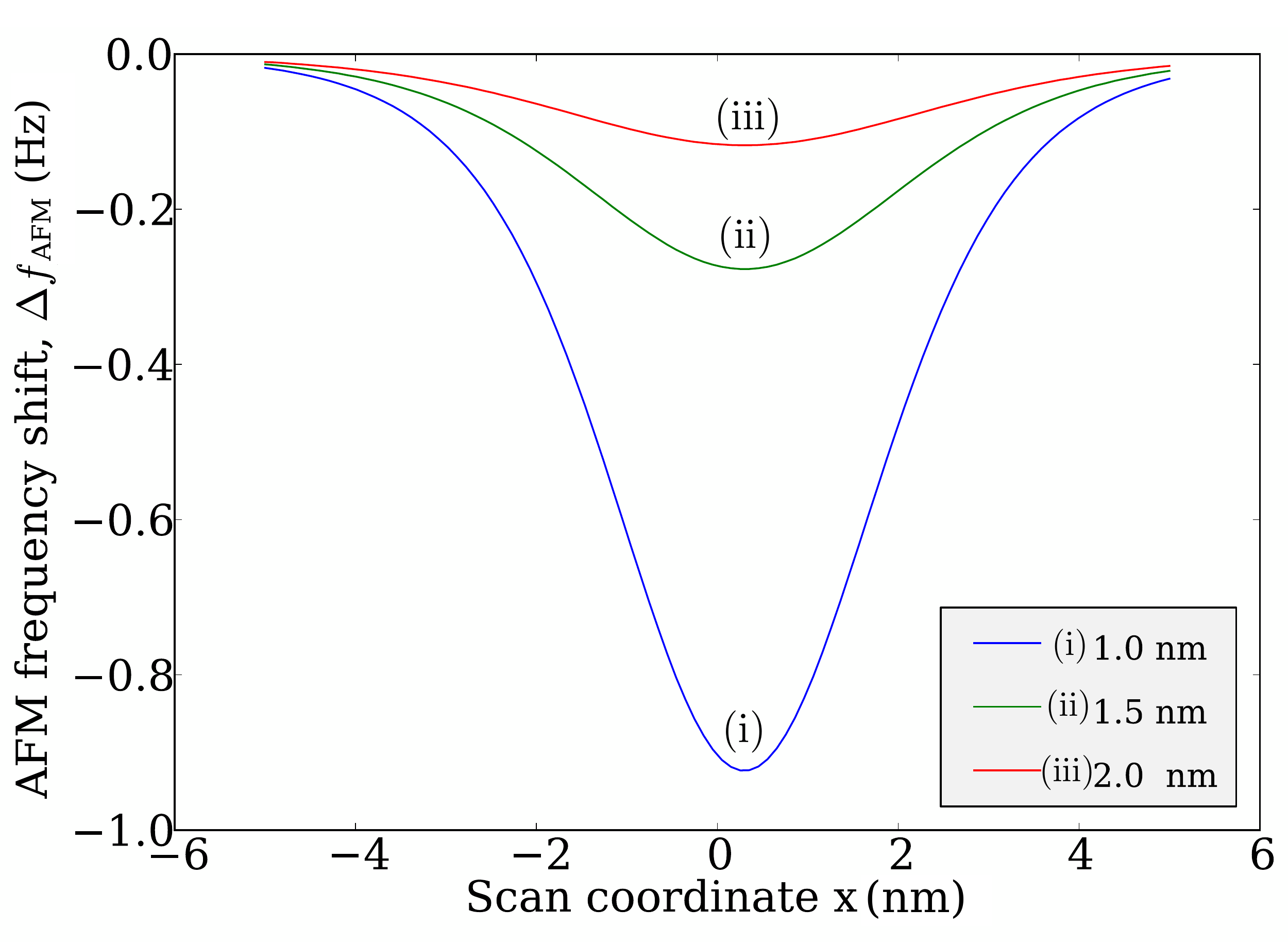}
\caption{AFM frequency shift as the tip scans along a line coinciding with
the DBP axis. The three curves correspond to as many equilibrium heights
of the AFM tip indicated in the legend, while the other scan
characteristics have the values indicated in the text.\label{fig:frequency-shift-vs-X}}
\end{figure}

Note that the task of sensing the location of a single charge as in
previous experimental work \cite{GML+09} is different from the current
task, where we attempt to obtain information about both the location
and the rates of (driven) motion of the electron. However, this does
not violate the uncertainty principle as we are not measuring the
$instantaneous$ location and momentum of the particle, but rather
time-averaged quantities, and the ultimate knowledge we aim to obtain
of the quantum system is statistical in nature.

In order to optimize the AFM read-out, we judiciously choose experimental
parameters. First, the AFM cantilever parameters
should be chosen so that the noise is much lower than the signal.
Second, for a given cantilever, we choose an appropriate oscillation amplitude
for the AFM tip. Larger amplitudes yield lower
noise, whereas lower amplitudes offer better spatial resolution. Finally,
$\xi$ should be maximized with respect to $x$ to achieve the largest possible frequency-shift read-out.

Although the greater sensitivity of silicon cantilevers is certainly
a desirable feature for the purpose of the current study, at present
it seems unlikely that they would allow the required atomic resolution
mainly due to their inability to achieve low-amplitude oscillations
and thus perform scans very close to the sample, 1-2 nm. Attempting such
tasks would likely lead to undesired frequent jump-to-contact events 
and thus very poor scans.

On the other hand, the qPlus tuning-fork system, although less sensitive,
has been already proven to
sense single charge with atomic spatial resolution~\cite{GML+09}
due to its robustness and ability to scan very close to the sample at low amplitudes.
(It easily avoids certain problems such as the jump-to-contact issue.\cite{Giessibl97}) 
It also allows combined AFM/STM studies thus facilitating DB fabrication and precise positioning
during the experiment. Therefore in this study we choose
parameters representative of the qPlus system, keeping in mind
that the optimal system may have characteristics somewhere 
in between those of the tuning fork and silicon cantilevers.

As experimental values, we henceforth assume $f_0= 30$~kHz, $k=1800$~N/m, $Q=5\times 10^4$,
$R_{\rm t}=$ 5~nm, $V_{\rm t}=$ 0~V,
and operation at liquid helium temperature. Also, unless otherwise
specified, the oscillation amplitude and equilibrium height for 
the AFM tip are assumed to be 3~\AA{}~and 1 nm, respectively.  
 The total noise in the frequency shift signal estimated for 
all the results below is less than 5~mHz at 4~K and 9~mHz at 77~K. As the AFM experiments yield the frequency shift in units of Hz, we present our results below in terms of $\Delta f_{\text{AFM}} = \Delta\omega_{\text{AFM}} / 2\pi$.
 
For a DBP$^{-}$ with separation of $7.68$~\AA{}, the AFM maximum
differential-frequency shift is obtained when the left dangling-bond
is $x\approx3$~\AA{}~away from the AFM tip central axis. 
Figure~\ref{fig:frequency-shift-vs-X} depicts the AFM frequency
shift $\Delta f_{\text{AFM}}$ as a function of the lateral tip position $x$. 
In this figure, it is clear that the effect of a trapped charge on the value of the AFM frequency is highly dependent on tip height. The great increase in signal for a tip height less than 1 nm 
is due to the fact that image-charge forces dominate in such close proximity to the localized charge.

Note that despite the simplicity of our model, 
the calculated magnitude of the signal is commensurate with past experimental results
of single-charge sensing with atomic resolution.\cite{GML+09}
Hence, this scheme is appropriately sensitive to small displacements of single
trapped charge.

\begin{figure}
	\includegraphics[width=\columnwidth]{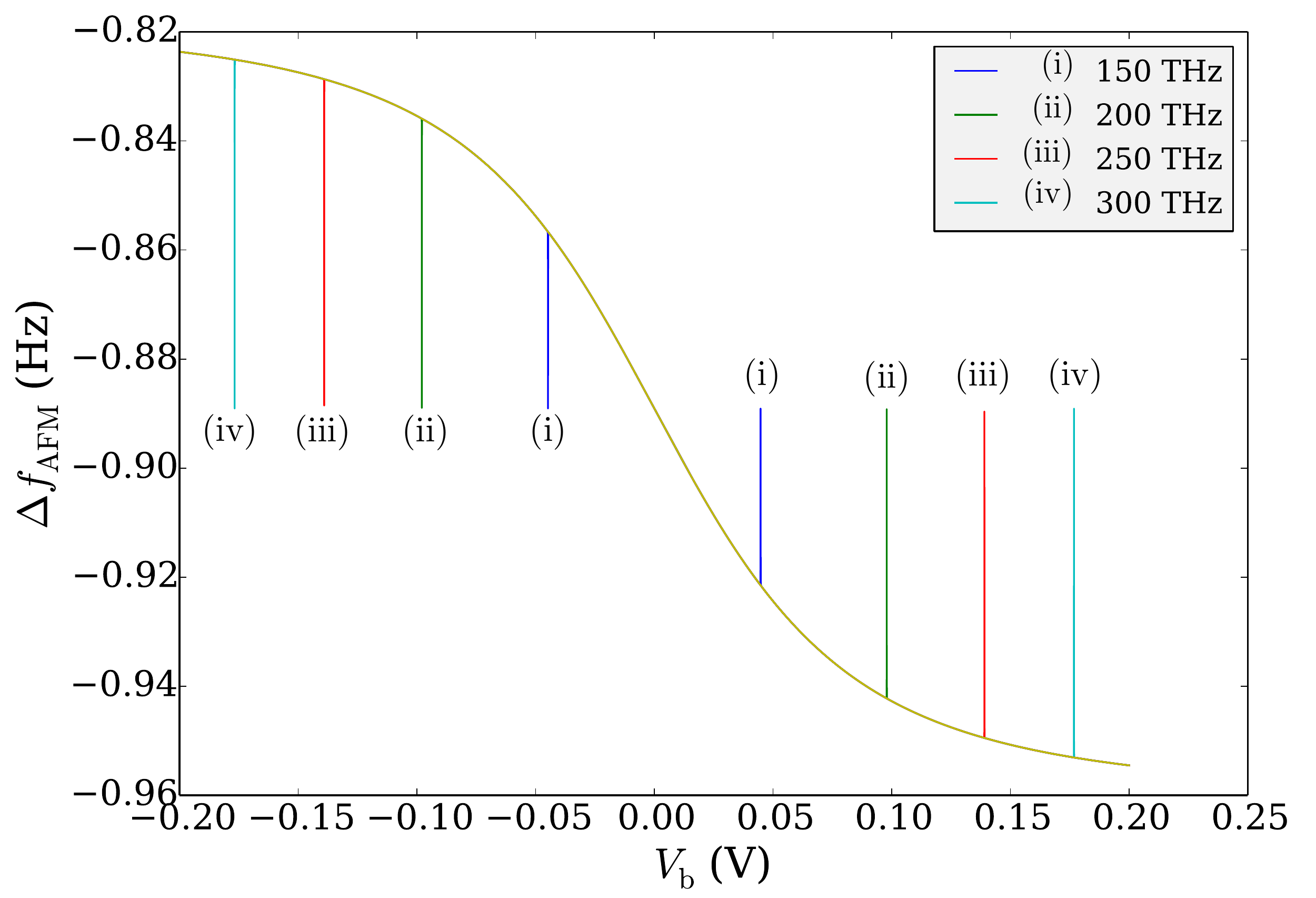}
\caption{AFM frequency shift ($\Delta f_{\text{AFM}}$) as a function of static bias (V$_\text{b}$) for four different
values of driving radiation frequency given in the legend and 
$\Omega_\text{MIR}$= 1 GHz.
At each frequency a set of two resonant peaks appear for two symmetric static 
bias values. 
\label{fig:frequency-shift-vs-Vb}}
\end{figure}

Figure~\ref{fig:frequency-shift-vs-Vb} shows the resonant peaks in the AFM signal. These resonant features are reflected in the oscillation frequency of the AFM tip when the DBP$^-$ is simultaneously exposed to a static bias and a driving radiation. In fact, the resonances can be exploited by varying the static bias for a fixed driving frequency. For each value of driving frequency a pair of resonant peaks appear on the AFM signal for two symmetric static-bias values.
These peaks contain information about our system and
can be used to determine the tunneling frequency $\Delta$ of the excess charge.

\begin{figure}
	\includegraphics[width=\columnwidth]{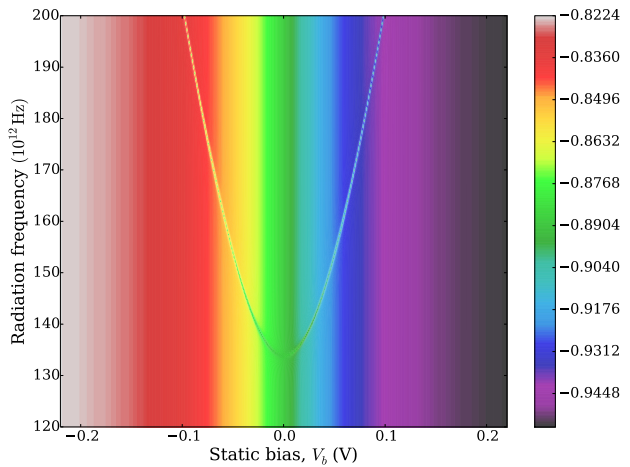}
\caption{Location of the resonance peaks in $\Delta f_{\mathrm{AFM}}$ in the
two-dimensional parameter space of the static bias and the driving
radiation frequency. The DBP$^-$ has a separation of 7.68~\AA~and 
the Rabi frequency is $\Omega_\text{MIR}$= 500 GHz.
\label{fig:Location-of-peaks2D}}
\end{figure}

More generally, 
Fig.~\ref{fig:Location-of-peaks2D} shows the AFM frequency shift
as we sweep both the MIR driving frequencies and the applied static bias
$V_{\text{b}}$ from a negative to a positive value. The resonance loci
appear here as ridges (trenches) when the MIR driving frequency is commensurate with the ramped
tunneling frequency. These resonance
trends mirror the parabolic relationship between the MIR driving frequency
and the static bias, as shown in Fig.~\ref{fig:2Dbias_vs_omegaNIR}.

\section{Damped dangling bond pair dynamics}
Until now our analysis ignores noise causing decoherence in the two-level system.
In this section we study the effect of noise on the shape and width of the resonances
used to characterize electron tunneling.
We show that the noise model can be tested by the measurements,
and we consider the specific model of spin-boson coupling~\cite{CL81}
to illustrate how the model is tested and the  parameters are acquired by measurement.

The spin-boson model characterizes weak coupling between a two-level system 
and a generic bosonic bath, such as phonons or charge fluctuations.\cite{CL81} 
It is described by the Hamiltonian
\begin{equation}
\label{eq:Hsb}
	\hat H_{\text{sb}}
		=\sum_i\hbar\omega_i \hat{b}_i^{\dagger}\hat{b}_i
			+ \hat\sigma_z \sum k_i\left(\hat{b}_i^{\dagger}+\hat{b}_i\right)
\end{equation}
with~$\omega_i$ the $i^\text{th}$ oscillator frequency,
$\hat{b}_i^{\dagger}$ and $\hat{b}_i$ the corresponding
creation and annihilation operators, and $k_i$ the coupling strength between 
the dangling-bond pair and the bath.

The first term on the right-hand side of Eq.~(\ref{eq:Hsb}) is the free bath Hamiltonian,
and the second term is the system-bath interaction Hamiltonian.
The interaction Hamiltonian indicates that the coupling depends linearly 
on the coordinates of the dangling-bond pair and the bath harmonic oscillators. 
The total Hamiltonian of the system would then be 
\begin{equation}
	\hat H=\hat H_{\text{b}}+\hat H_{\text{d}}+\hat H_{sb},
\end{equation}
where the first two terms are given in Eqs.~(\ref{eq:Hsystem}) and~(\ref{Hdrive}). 

Solving the Lindblad master equation for the dangling-bond pair,\cite{Lin76}
the steady-state probability for the excess charge to be in the left quantum dot is~\cite{BS06, SDB05}
\begin{equation}
\label{eq:rho}
	\rho_{\text{L-sb}}
		=\frac{1}{2}
			+\frac{eV_{\text{b}}\Gamma_{\text{r}}}{2\hbar\Delta'}
			\frac{\eta^2+\Gamma_{\phi'}^2}
			{\eta^2\Gamma_{\text{r}}+\Omega_{\text{MIR}}^2\Gamma_{\phi'}
			+\Gamma_{\text{r}}\Gamma_{\phi'}^2},
\end{equation}
with decoherence rate $\Gamma_{\phi'}=\Gamma_{\phi}+\Gamma_{\text{r}}/2$
for relaxation rate~$\Gamma_{\text{r}}$ and dephasing rate~$\Gamma_{\phi}$.
For $\Gamma_{\phi'}\rightarrow 0$ and $\tan\varphi=\frac{|\Omega|}{\eta}$,
Eq.~(\ref{eq:rho}) reduces to Eq.~(\ref{eq:rho vs bias vs driven}).
In another limit,
the relaxation rate $\Gamma_{\text{r}}$ and dephasing rate $\Gamma_{\phi}$ 
are equal up to second order in the limit of weak qubit-bath coupling:
$\Gamma:=\Gamma_r\approx\Gamma_\phi$.
%

\begin{figure}
	\includegraphics[width=\columnwidth]{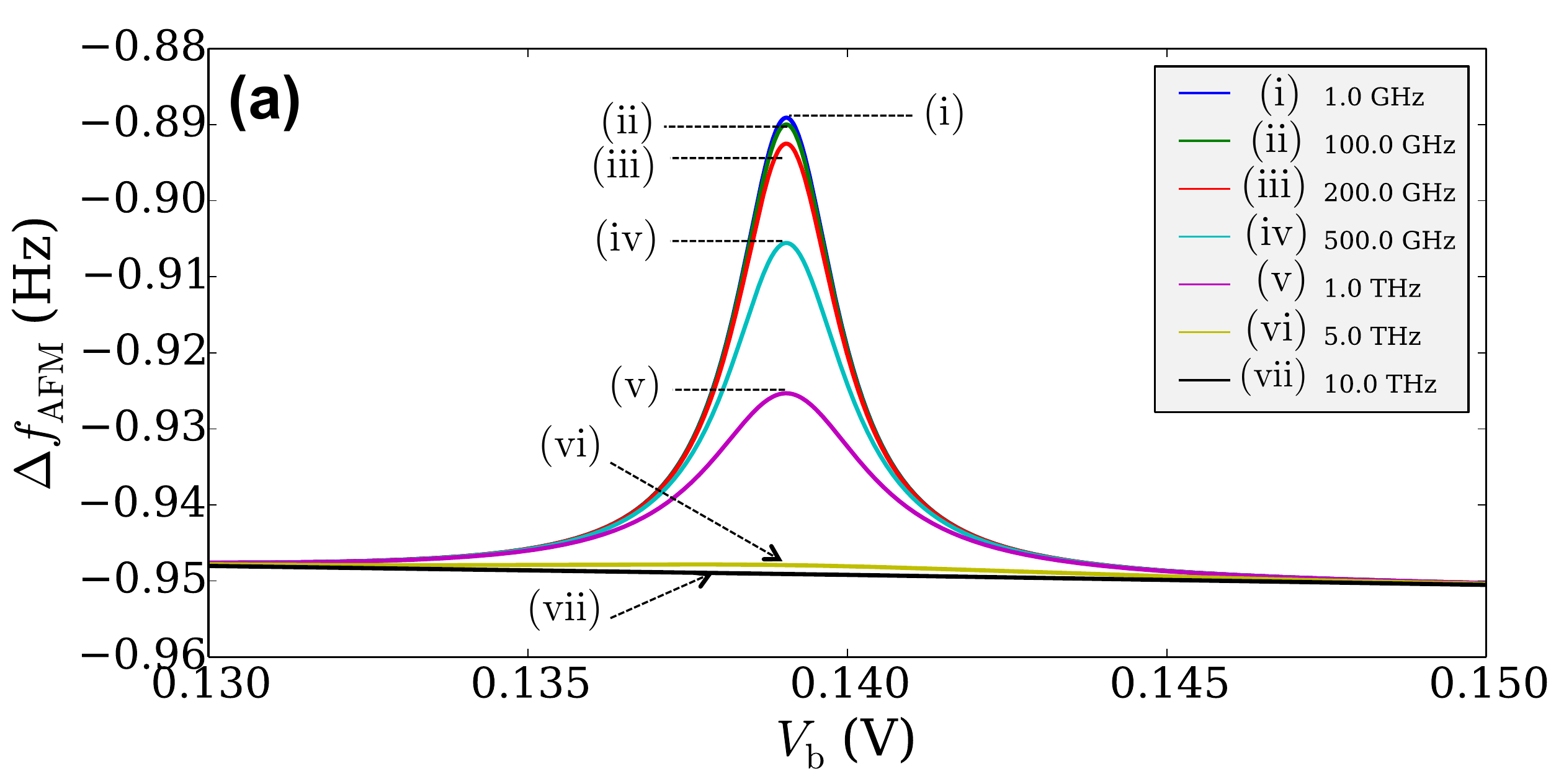} \\
	\includegraphics[width=\columnwidth]{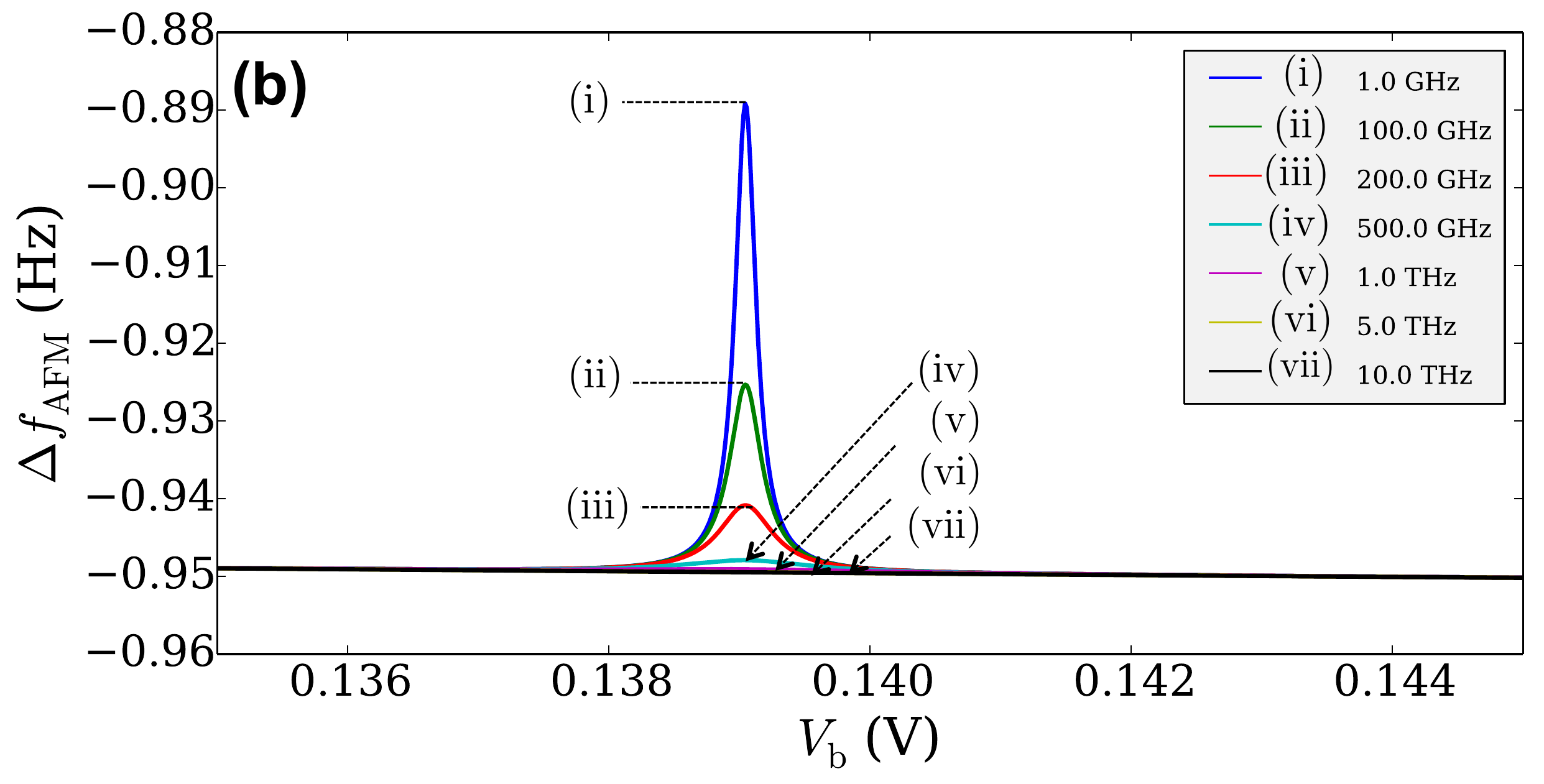} \\
	\includegraphics[width=\columnwidth]{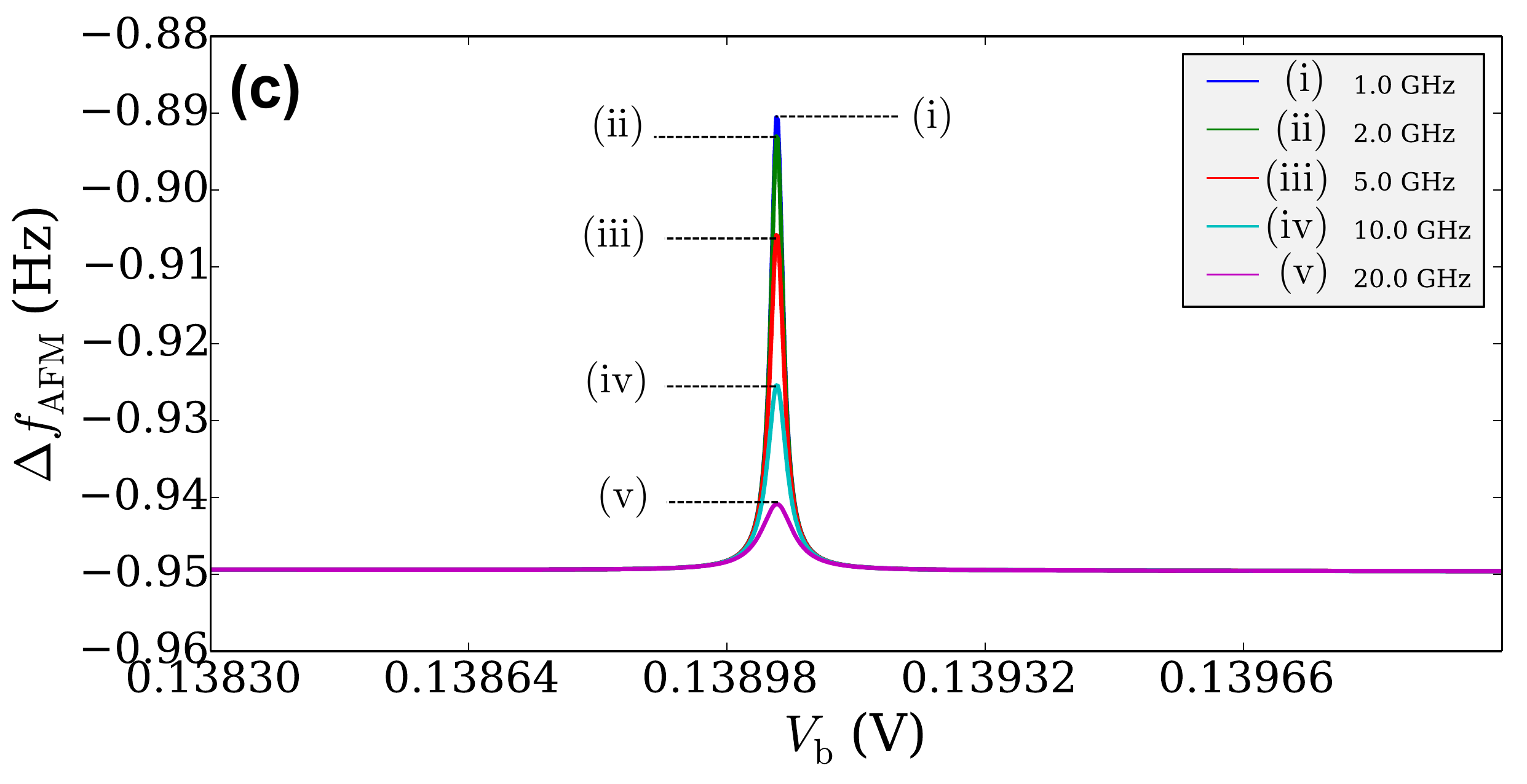} \\
\caption{
AFM frequency shift as a function of static bias for different decoherence rates shown in the legends and for three chosen values of the Rabi frequency: 1 THz in (a), 100 GHz in (b), and 10 GHz in~(c). The laser frequency was fixed to $\omega_{\text{MIR}}$= 250 THz in all cases. Note the progressive narrowing of the range on the horizontal axis from top to the bottom panels.
}
\label{fig:decohering-freq-vs-Vb}
\end{figure}

The AFM frequency shift is shown in Fig.~\ref{fig:decohering-freq-vs-Vb} as a function of static 
bias for MIR driving frequency $\omega_{\text{MIR}}$
fixed to 250~THz and various decoherence rates given in the legends.
As the Rabi frequency is sampled in decreasing order over three orders of magnitude in (a), (b), and (c), 
we notice strong narrowing of the resonance peaks,also seen in Fig.~\ref{fig:laser_intensity_effect}
in the absence of any decoherence. 

A different range of the decoherence rate was sampled in each case in order to capture 
the main effect: peak height decreases with increasing decoherence rate and the effect
is measurable when the Rabi frequency is commensurate (same order of magnitude) with the decoherence rate. This is similar to the behavior of a critically damped driven harmonic oscillator.

For a Rabi frequency exactly equal to the decoherence rate, the peak height is about 40\% of its
predicted decoherence-free value. Note that the plots in (c) are predicted to be applicable 
at low temperature (4K), while (a) and (b) could be used at higher temperatures
if the decoherence rates go up. As our proposed experiment is to take place at 
temperatures of 4~K and higher, 
a Rabi frequency of 10 GHz seems sufficient to capture these decoherence signatures
in the low temperature regime. 
This corresponds to an applied laser intensity of about 2$\times 10^5$ W/m$^2$.
Note however that an even smaller laser intensity (2$\times 10^3$ W/m$^2$) is sufficient
if one does not seek decoherence measurements, but only the native tunneling rates 
of the qubit.

Experimentally, the limiting factors in measuring these peaks are the (horizontal) resolution 
in the voltage on the biasing electrodes, and the (vertical) resolution in the frequency shift
of the AFM. 
In practice, measured resonant peaks can be compared with model curves 
to yield information about the decoherence rates.
For a sufficiently weak MIR field, the full-width-at-half-maximum height of each peak
directly yields the decoherence rate.
Even if the coupling strength between the MIR field and the DBP$^-$ is unknown,
the decoherence rate can still be determined by extracting the resonant peak height as a function of 
width for various values of the power of the incident MIR field.\cite{BS06}

\begin{figure}
	\includegraphics[width=\columnwidth]{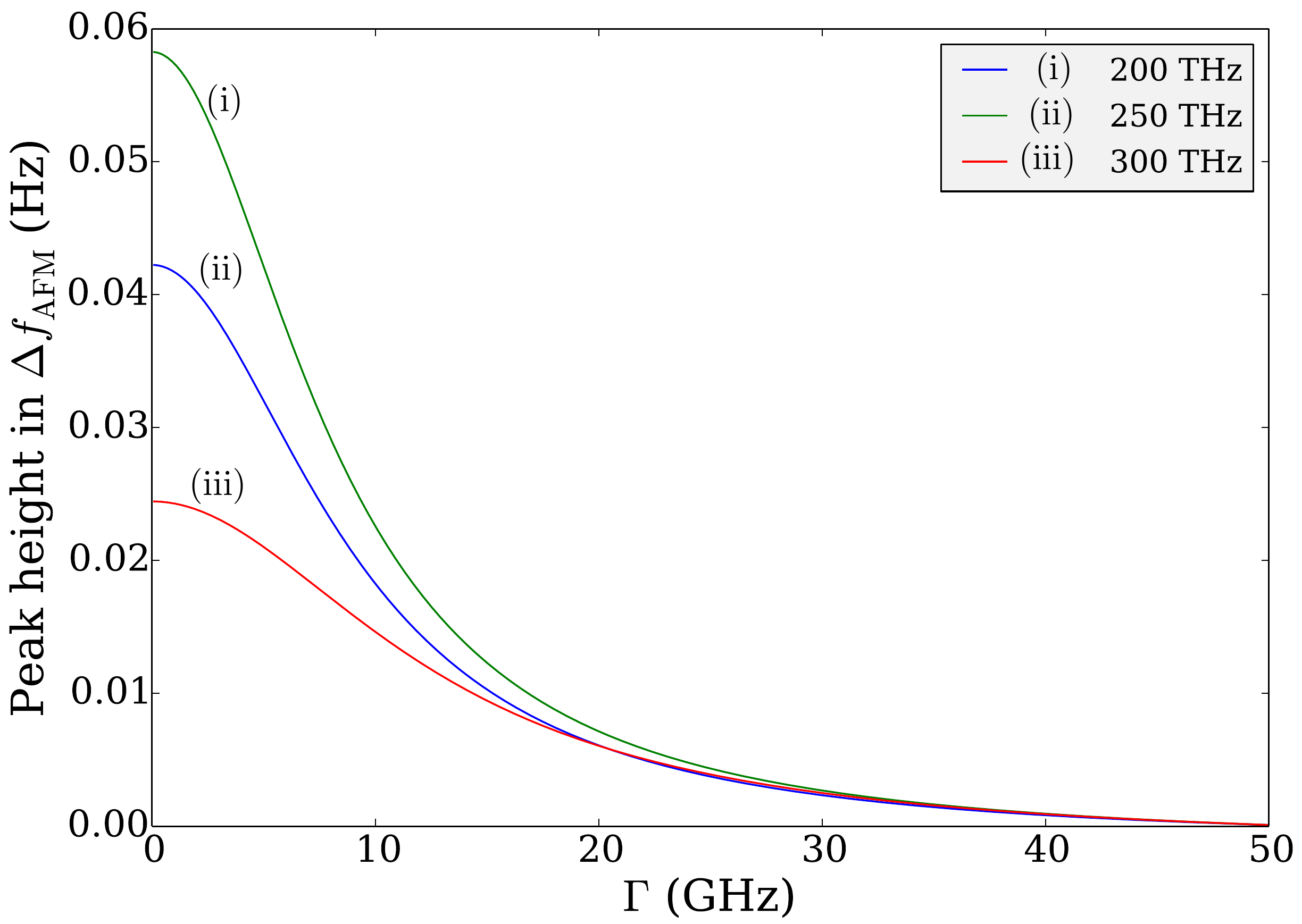}
\caption{
	Resonant peak magnitudes in the AFM frequency shift 
as a function of decoherence rate $\Gamma$ for different fixed frequency $\omega_{\text{MIR}}$ of the driving radiation shown in the legend and 
for a Rabi frequency set at 10 GHz.
	}
\label{fig:peaks-vs-gamma}
\end{figure}

Fig.~\ref{fig:decohering-freq-vs-Vb}
conveys three key points of our proposal:
how the tunneling rate $\Delta$ can be inferred, how the decoherence model can be tested,
and how the decoherence rate $\Gamma$ is obtained if the model is correct.
The tunneling rate is revealed by observing resonances of the AFM frequency shift
and is obtained by choosing~$\omega_{\text{MIR}}$ and $V_{\text{b}}$ judiciously.
The decoherence model is tested by seeing whether the frequency shift obeys the model-predicted 
dependence on driving-field frequency and static bias.
Finally, the decoherence rate is obtained by comparing the measured resonance peak heights
with those predicted for a decoherence-free system. 
The plots in Fig.~\ref{fig:peaks-vs-gamma} serve as examples of expected 
behavior and can be used in practice for extracting decoherence parameters from experimental data.

In the spin-boson decoherence model discussed here,
the decoherence rate can be determined solely from the width of the peak
because $\Gamma_{\text{r}}$ and $\Gamma_{\phi}$ 
are equal up to second order in the limit of weak qubit-bath coupling.
For a noise model with independent relaxation and dephasing rates,
Eq.~(\ref{eq:rho}) demonstrates sensitivity to changes in each of these rates independently.

A practical concern in performing the experiment is the back-action
of our detector (AFM) on the quantum system being measured. 
The AFM cantilever has a frequency of $10^4$ Hz, whereas 
the oscillation frequency of the dangling-bond excess-charge is estimated to be $10^{14}$ Hz.
Thus, the AFM tip motion looks adiabatic to the excess charge and 
the back-action of the tip on coherent electron dynamics in the dangling-bond pair is negligible.

There is, however, a static component of the tip perturbation on the qubit, 
which is the flipside of the AFM sensitivity 
of charge location: the tip creates a static bias along the DBP$^-$ axis.
We calculated this bias for our optimized setup to have a maximum of 15.8 meV
at the spatial location of maximum AFM sensitivity. 
Fortunately, this static bias does not significantly alter our scheme as it just
adds to the applied static bias, and the resonance peaks will still be obtained
albeit with a horizontal shift of 15.8 meV. Therefore, correcting for this back-action
is just a simple matter of recalibrating
the horizontal ($V_{\rm b}$) axis in our plots so the peak locations become again symmetric.
Thus we have the ability to measure this shift 
and compensate for it in any subsequent data processing.

\section{Summary and conclusion}
We have proposed a feasible experimental scheme for characterizing the fast 
tunneling rate as well as the nature and rate of decoherence of an excess 
charge shared between a pair of coupled dangling-bonds on the surface of silicon.
In our scheme, the electrostatic potential across the dangling-bond pair is ramped by external electrodes. 
Furthermore the dangling-bond pair is driven by a MIR field,
and the resulting resonances correspond to equal distribution of the electron location 
in the dangling-bond pair despite, and independent of, the strength of 
the static bias thereby revealing the desired tunneling properties.

The distribution of the excess electron between left and right dangling bonds is 
detected by capacitively coupling the DB excess charge to an atomic force microscope tip.
Resonances are observed on the AFM frequency-shift signal when the 
MIR field matches the ramped tunneling frequency of the excess charge. 

Experimentally, charge qubit geometries must be chosen so that tunnel splittings (and corresponding driving frequencies)
should avoid undesired excitations such as the different vibration modes of H-Si bonds\cite{LNK79, Luc79}  in the interval from
526 to~1111 cm$^{-1}$.
In practice, a control experiment would first be used to calibrate the AFM probe in 
the absence of driving radiation.
In order to calibrate the vertical oscillation frequency as a function of lateral position, 
an AFM tip will be placed at different positions near a \emph{single} DB$^{-}$. 
The AFM vertical oscillation frequency will exhibit a shift that depends on the lateral position
of the tip with respect to the charge and the tip height. Oscillation amplitudes and tip height
can then be adjusted to obtain maximum signal.

Our scheme will enable in-depth studies of quantum coherent transport of electrons between dangling
bonds on the surface of silicon and enable the study of phonons and other interactions.
As dangling bond systems are promising building blocks for quantum-level engineering of novel devices
including quantum-dot cellular automata\cite{HPD+09} and quantum computing,\cite{LXS+10}
a detailed quantitative analysis of electron dynamics in dangling bond assemblies is an important step.

\acknowledgments
We greatly appreciate valuable discussions with S. D. Barrett and T. M. Stace. This project has been supported by NSERC, AITF, and CIFAR.

\bibliography{danglingbondpairs}

\end{document}